\documentclass[a4paper,11pt]{article}

\pdfoutput=1 

\usepackage{jcappub} 

\usepackage[T1]{fontenc} 

\usepackage{subfig}
\usepackage{graphicx}
\usepackage{amsmath}
\usepackage{mathtools}
\usepackage{nicefrac}
\usepackage[dvipsnames]{xcolor}
\usepackage{booktabs}
\usepackage[normalem]{ulem}
\usepackage{physics}

\newcommand{\textcode}[1]{{\large\scshape#1}}

\newcommand{\Msun}{\, h^{-1} \,  M_{\odot}}
\newcommand{\hompc}{\,h\,{\rm Mpc}^{-1}}
\newcommand{\mpcoh}{\,h^{-1}\,{\rm Mpc}}

\renewcommand{\vec}{\mathbf}
\newcommand{\vbar}{\Bar{v}}
\NewDocumentCommand{\codeword}{v}{%
\texttt{#1}}
\newcommand{\matr}[1]{\mathbf{#1}}

\title{\boldmath Studying large-scale structure probes of modified gravity with COLA}
\author[a,1]{Bartolomeo Fiorini,\note{Corresponding author.}}
\author[a]{Kazuya Koyama,}
\author[a]{Albert Izard}

\affiliation[a]{Institute of Cosmology \& Gravitation, University of Portsmouth, Dennis Sciama Building, Burnaby Road, Portsmouth, PO1 3FX, United Kingdom}

\emailAdd{bartolomeo.fiorini@port.ac.uk}

\abstract{We study the effect of two Modified Gravity (MG) theories, $f(R)$ and nDGP, on three probes of large-scale structure, the real space power spectrum estimator $Q_0$, bispectrum and voids, and validate fast approximate COLA simulations against full {\it N}-body simulations for the prediction of these probes. %
We find that using the first three even multipoles of the redshift space power spectrum to estimate $Q_0$ is enough to reproduce the MG boost factors of the real space power spectrum for both halo and galaxy catalogues. %
By analysing the bispectrum and reduced bispectrum of Dark Matter (DM), we show that the strong MG signal present in the DM bispectrum is mainly due to the enhanced power spectrum. We warn about adopting screening approximations in simulations as this neglects non-linear contributions that can source a significant component of the MG bispectrum signal at the DM level, but we argue that this is not a problem for the bispectrum of galaxies in redshift space where the signal is dominated by the non-linear galaxy bias. %
Finally, we search for voids in our mock galaxy catalogues using the \texttt{ZOBOV} watershed algorithm.
To apply a linear model for Redshift-Space Distortion (RSD) in the void-galaxy cross-correlation function, we first examine the effects of MG on the void profiles entering into the RSD model. We find relevant MG signals in the integrated-density, velocity dispersion and radial velocity profiles in the nDGP theory. Fitting the RSD model for the linear growth rate, we recover the linear theory prediction in an nDGP model, which is larger than the $\Lambda$CDM prediction
at the $3 \sigma$ level. In $f(R)$ theory we cannot naively compare the results of the fit with the linear theory prediction as this is scale-dependent, but we obtain results that are consistent with the $\Lambda$CDM prediction.}

\begin{document}
\maketitle
\flushbottom

\section{Introduction}
Thanks to upcoming stage IV experiments such as the space-based Euclid mission \cite{Laureijs11} and the ground-based Rubin observatory's Legacy Survey of Space and Time (LSST) \cite{lsst09} and the Dark Energy Spectroscopic Instrument (DESI) \cite{DESI:2018ymu}, we are entering an era of unprecedented detailed cosmological observations of the Large-Scale Structure (LSS). Joint multi-probe techniques are a promising direction to break the degeneracies in the theory parameter space, enabling tight constraints on cosmological parameters at intermediate redshift. In particular, since structure formation is tightly linked to the dynamics of gravity, it is key to detect any deviation from General Relativity (GR) in future analyses, so that we better understand gravity on cosmological scales and also to get unbiased inferences of cosmological parameters.

The simplest extensions to GR can be described by an extra degree of freedom in the form of a scalar field that mediates an additional force, the so-called fifth force. The stringent Solar System tests of gravity impose tight constraints on the strength of the fifth force \cite{Will:2014kxa} which needs to be weak in such environments, therefore Modified Gravity (MG) theories of cosmological interest incorporate screening mechanisms that suppress the effect of the fifth force in high-density regions \cite{Koyama:2015vza,Joyce:2014kja}. 

Cosmological simulations are indispensable to create model predictions on non-linear scales and to better constrain the relationship between galaxies and the underlying Dark Matter (DM) density field, encapsulated in the galaxy bias. The non-linear regime is also where the screening mechanisms can leave their imprint. Full {\it N}-body simulations enable accurate predictions of the DM density field on non-linear scales but they are computationally very expensive. Their extension to MG are even more expensive because they have to solve the non-linear differential equation for the extra scalar field \cite{Winther:2015wla}.

Approximate simulation methods are computationally cheaper alternatives to full {\it N}-body simulations. In particular, the COmoving Lagrangian Acceleration (COLA) method \cite{Tassev:2013pn} combines second-order Lagrangian perturbation theory (2LPT) to model analytically the trajectories of DM particles on large scales, with an {\it N}-body method known as the Particle Mesh (PM) algorithm to take care of small scale physics. This separation between large scales and small scales allows COLA simulations to trade accuracy on small scales for a great reduction in computational cost while leaving the accuracy on large scales unaffected. The extension of the COLA method to MG theories requires the use of screening approximations to avoid a significant impact on their computational efficiency. Thanks to the use of screening approximations based on spherically symmetric solutions developed in  \cite{Winther:2014cia}, the COLA method was extended to MG theories while still being computationally efficient \cite{Valogiannis:2016ane, Winther:2017jof}. Despite the limited accuracy of COLA in the deep non-linear regime, it has been shown to have a good performance in predicting the reaction of the matter power spectrum to changes in cosmological parameters, with sub-percent accuracy up to $k\sim 1 \hompc$ for a wide range of cosmologies around the reference one \cite{Brando:2022gvg}.

Several empirical prescriptions have been proposed in the literature to link the DM density field predicted by DM-only cosmological simulations to the distribution of galaxies. Among them, the widely used Halo Occupation Distribution (HOD) prescription has been extended and validated to produce galaxy mocks catalogues from COLA simulations in $f(R)$ \cite{Starobinsky:1980te} and nDGP \cite{Dvali:2000hr} gravity theories, reproducing full {\it N}-body results for the MG signals on the redshift-space multipoles of the galaxy power spectrum \cite{Fiorini:2021dzs}. 

Given the large number of parameters that enter the modelling of galaxy clustering (e.g. cosmological parameters, MG parameters, HOD parameters), degeneracies in the high-dimensional theory parameter space can affect the constraining power of the LSS. However, the theory parameters may have different impacts on different summary statistics. An example of this is the degeneracy between massive neutrinos and $\sigma_8$ in the power spectrum which is broken with bispectrum information \cite{Hahn:2019zob,Hahn:2020lou}. 

In this work, we exploit the results and simulated data sets from \cite{Fiorini:2021dzs} to investigate the impact of MG in statistics beyond the redshift-space multipoles of the power spectrum, and by doing so we validate the COLA method to formulate theoretical predictions on these additional statistics in MG theories. 
As mentioned, COLA simulations developed in \cite{Winther:2017jof, Wright:2017dkw} use a screening approximation that linearises the scalar field equation while including an environmental dependent screening factor. The validity of this approximation needs to be checked for higher-order statistics as they lack contributions from the non-linearity of the scalar field. 
We focus on two MG theories, the Hu-Sawicky $f(R)$ theory \cite{Hu:2007nk} for $n=1$ and the normal branch of the 5D braneworld theory known as Dvali-Gabadadze-Porrati gravity (DGP) \cite{Dvali:2000hr}, because they are good representatives of the screening mechanisms they incorporate, chameleon \cite{Khoury:2003aq, Khoury:2003rn} and Vainshtein mechanism \cite{Vainshtein:1972sx}, respectively. After reviewing the simulations and catalogues at the base of this analysis, we test the effects of MG theories and assess COLA's accuracy on the power spectrum orthogonal to the line-of-sight in section~\ref{sec:PowerSpectrum}, on the bispectrum in section~\ref{sec:Bispectrum} and on voids in section~\ref{sec:Voids}. We then conclude in section~\ref{sec:conclusion}.

\section{Simulations and HOD tuning}
Theoretical predictions on the redshift-space power spectrum (and higher-order statistics) of galaxies can be formulated from mock galaxy catalogues produced using cosmological simulations with the HOD prescription. 

In this work, we use data from $z\simeq0.5$ of the PITER and ELEPHANT simulations suites produced in \cite{Fiorini:2021dzs} for the extension and validation of the pipeline for the production of galaxy mock catalogues in MG with the COLA method. The ELEPHANT simulations \cite{Cautun:2017tkc} are full {\it N}-body simulations in GR, $f(R)$, and nDGP gravity computed with the \codeword{ECOSMOG} code \cite{Li:2011vk}, which implements the adaptive mesh-refinement technique and solves the equation for the fifth force without approximations. The PITER simulation suite introduced in \cite{Fiorini:2021dzs} is a set of COLA simulations designed to mimic the ELEPHANT simulations. Both PITER and ELEPHANT suites (COLA and {\it N}-body respectively from now on) follow the evolution of $1024^3$ particles in a $(1024 \mpcoh)^3$ box and consist of 5 realisations in GR, F5 and N1 gravity\footnote{In each realisation the same initial seed is used for all gravity models, but COLA and {\it N}-body simulations have different initial seeds.}. F5 refers to the Hu-Sawicki $f(R)$ model with $n=1$ and  $|f_{R0}| = 10^{-5}$ and N1 to the normal branch of the DGP model (nDGP) with $H_0 r_c=1$. The cosmology characterising these simulations is 
\begin{equation}
\begin{array}{ccc}
\Omega_{m,0}=0.281 \, , & \Omega_{\Lambda,0}=0.719 \, , & \Omega_{b,0}=0.046 \, , \\
n_{s}=0.971 \, , & \sigma_{8}=0.842 \, , & h=0.697 \, ,
\end{array}
\label{cosmology}
\end{equation}
and additional technical details are summarised in table~\ref{tab:piter_details} and table~\ref{tab:elephant_details} (see \cite{Fiorini:2021dzs} for a comparison of the settings adopted in the two simulations suites and \cite{Fiorini:2021dzs,Cautun:2017tkc} for more details on the simulations settings).

\begin{table}
\centering 
\parbox{.45\linewidth}{
\caption{\textcode{piter} simulations} \label{tab:piter_details}
\begin{tabular}{ccccc}
\toprule
Models & GR, F5, N1 \\
Realisations & 5 \\
Box size & 1024 $\mpcoh$ \\
$N_{\mathrm{part}}$ & $1024^3$ \\
$M_{\rm part}$ & $7.7985 \cdot 10^{10}\Msun$ \\
Force grid & $3072^3$ \\
Timesteps & $30$ \\
Initial conditions & 2LPT, $z=49$ \\
\bottomrule
\end{tabular}
}
\hfill
\parbox{.45\linewidth}{
\caption{\textcode{elephant} simulations} \label{tab:elephant_details}
\begin{tabular}{cc}
\toprule
Models & GR, F5, N1 \\
Realisations & 5 \\
Box size & 1024 $\mpcoh$ \\
$N_{\mathrm{part}}$ & $1024^3$ \\
$M_{\rm part}$ & $7.7985 \cdot 10^{10}\Msun$ \\
Domain grid & $1024^3$ \\
Refinement criterion & $8$ \\
Initial conditions & Zel'dovich, $z=49$ \\
\bottomrule
\end{tabular}
}
\end{table}

The halo catalogues were created from the snapshots of COLA and {\it N}-body simulations at redshift $z=0.5057$ using a friend-of-friend halo finder with linking length of $0.2$ in units of the mean inter-particle distance.  The halo catalogues were populated with galaxies by means of the HOD model proposed in \cite{Zheng:2007zg} for the occupation numbers of central and satellite galaxies. To enable the accurate modelling of MG effects on the halo profiles and consequently on the phase-space distribution of satellite galaxies, in F5 theory, the concentration-mass relation and the velocity-dispersion profile for the Navarro-Frenk-White model \cite{Navarro:1995iw} were modified based on fitting functions calibrated on {\it N}-body halo catalogues produced with the \codeword{Rockstar} halo-finder \cite{Behroozi13}. The HOD tuning was performed in MG theories and in COLA simulations to reproduce the monopole $P_0$ and the quadrupole $P_2$ of the galaxy power spectrum\footnote{See section~\ref{sec:PowerSpectrum} for the definition of the power spectrum and its multipoles in redshift space.} in redshift-space up to $k_{\rm max} = 0.3 \hompc$ 
estimated using {\it N}-body simulations in GR with the HOD parameters in eq.~(26) of \cite{Manera:2012sc}, originally obtained to reproduce the clustering of CMASS galaxies in the data release 9 of the BOSS survey. 

All the halo samples drawn in this work follow the same abundance matching criteria as in \cite{Fiorini:2021dzs}. That is, for each simulation, the most massive halos are selected while matching the reference abundance determined in the {\it N}-body catalogues in GR using a mass-cut at $M_{\rm cut} = 10^{13} \, \Msun$.

To increase the signal-to-noise ratio in summary statistics at the galaxy level, \cite{Fiorini:2021dzs} relied on a set of 5 HOD realisations produced from each simulation realisation and, in addition to this, redshift-space statistics were estimated by averaging over the 3 lines-of-sight parallel to the box's major axes. We adopt the same technique for the galaxy statistics discussed in this work.

\section{Power spectrum}
\label{sec:PowerSpectrum}
The real space power spectrum
\begin{equation}
    \left\langle\delta\left(\vec{k}\right) \delta\left(\vec{k}'\right)\right\rangle \equiv(2 \pi)^{3} P^{r}\left(k\right) \delta_{D}\left(\vec{k}+\vec{k}'\right) \,.
\end{equation}
is not directly measurable in galaxy surveys because we cannot probe the real space position of galaxies. What we can directly measure is the redshift-space power spectrum $P^{s}(k, \mu)$ where $\mu \equiv \hat{z}\cdot \hat{k}$ and $\hat{z}$ is the line-of-sight in the plane-parallel approximation. The redshift-space power spectrum can be projected on the orthonormal basis of Legendre polynomials
\begin{equation}\label{multipoles}
    P_{\ell} (k) = \frac{2\ell +1}{2} \int_{-1}^{1}P^{s}(k, \mu) \mathcal{L}_{\ell}(\mu) d\mu \, ,
\end{equation}
where $\mathcal{L}_{\ell}$ is the Legendre polynomial of order $\ell$, obtaining the multipole of the power spectrum $P_{\ell}$.

\subsection{Estimator of the real space power spectrum}
\label{sec:Q0}
Before studying statistics beyond the power spectrum, we first study the method to estimate the real space power spectrum from the combination of redshift-space multipoles proposed in \cite{Ivanov:2021fbu}. 

From the definition of the power spectrum multipoles and given the orthogonality of the Legendre polynomials, the redshift-space power spectrum $P^s(k, \mu)$ can be reconstructed using the multipoles $P_{\ell}$
\begin{equation}\label{Ps_LagrangeExp}
    P^s(k, \mu)=\sum_{\ell} P_{\ell}(k) \mathcal{L}_{\ell}(\mu) \, .
\end{equation}
On the other hand, $P^s(k, \mu)$ can also be expanded in the monomial basis as
\begin{equation}
    P^s(k, \mu)=\sum_{n} Q_{n}(k) \mu^n \, .
\end{equation}
An interesting quantity arising from the latter expansion is the first moment $Q_0 = P^s(k, 0)$ which, representing the power spectrum orthogonal to the line-of-sight, is unaffected by Redshift-Space Distortions (RSD) and coincides with the real space power spectrum $P^{r}\left(k\right)$. Using Eq.\eqref{Ps_LagrangeExp}, $Q_0$ can be expressed in term of the power spectrum multipoles $P_{\ell}$
\begin{equation}\label{Q0ofPell}
    Q_0 = \sum_{\ell} P_{\ell}(k) \mathcal{L}_{\ell}(0) \, .
\end{equation}
The Kaiser model for RSD \cite{Kaiser:1987qv} predicts that linear velocity inflow towards over-densities produces power transfer from the real space power spectrum to the quadrupole and hexadecapole of the power spectrum in redshift-space
\begin{equation}
   \delta^{s}(\vec{k})=\left[1+f \mu^{2}\right] \delta^{r}(\vec{k}) \, .
\end{equation}
On top of this, the Fingers of God (FoG) effect \cite{Jackson:1971sky}, due to the random non-linear motion of galaxies on small scales, further transfers power to higher multipoles. When the FoG effect can be neglected, the sum in Eq.~\eqref{Q0ofPell} can be restricted to the multipoles sourced by the Kaiser effect \cite{Scoccimarro:2015bla,Ivanov:2021fbu}
\begin{equation}\label{Q0ofP024}
    Q_0 = \sum_{\ell=0,2,4} P_{\ell}(k) \mathcal{L}(0) = P_{0}-\frac{1}{2} P_{2}+\frac{3}{8} P_{4} \, .
\end{equation}
Otherwise higher multipoles should be considered in the expansion, and since these are more affected by noise than lower multipoles, approaches such as the one of \cite{Ivanov:2021fbu} can be used to mitigate the loss of precision.

Since $Q_0$ is unaffected by RSD, theoretical modelling is easier than for the redshift-space multipoles $P_{\ell}$, and this fact was used to extend the validity range of perturbation theory-based approaches such as EFTofLSS \cite{Ivanov:2021fbu}. We limit our analysis to the $Q_0$ estimate by means of the truncated sum of Eq.~\eqref{Q0ofP024}, which we measure in our halo and galaxy catalogues and compare to the real space power spectrum to gain insight into the effect of MG on this observable and into the validity of COLA simulations to make theoretical predictions for $Q_0$.

\subsection{Results for power spectrum}
\begin{figure}
\centering
\includegraphics[width=.98\textwidth]{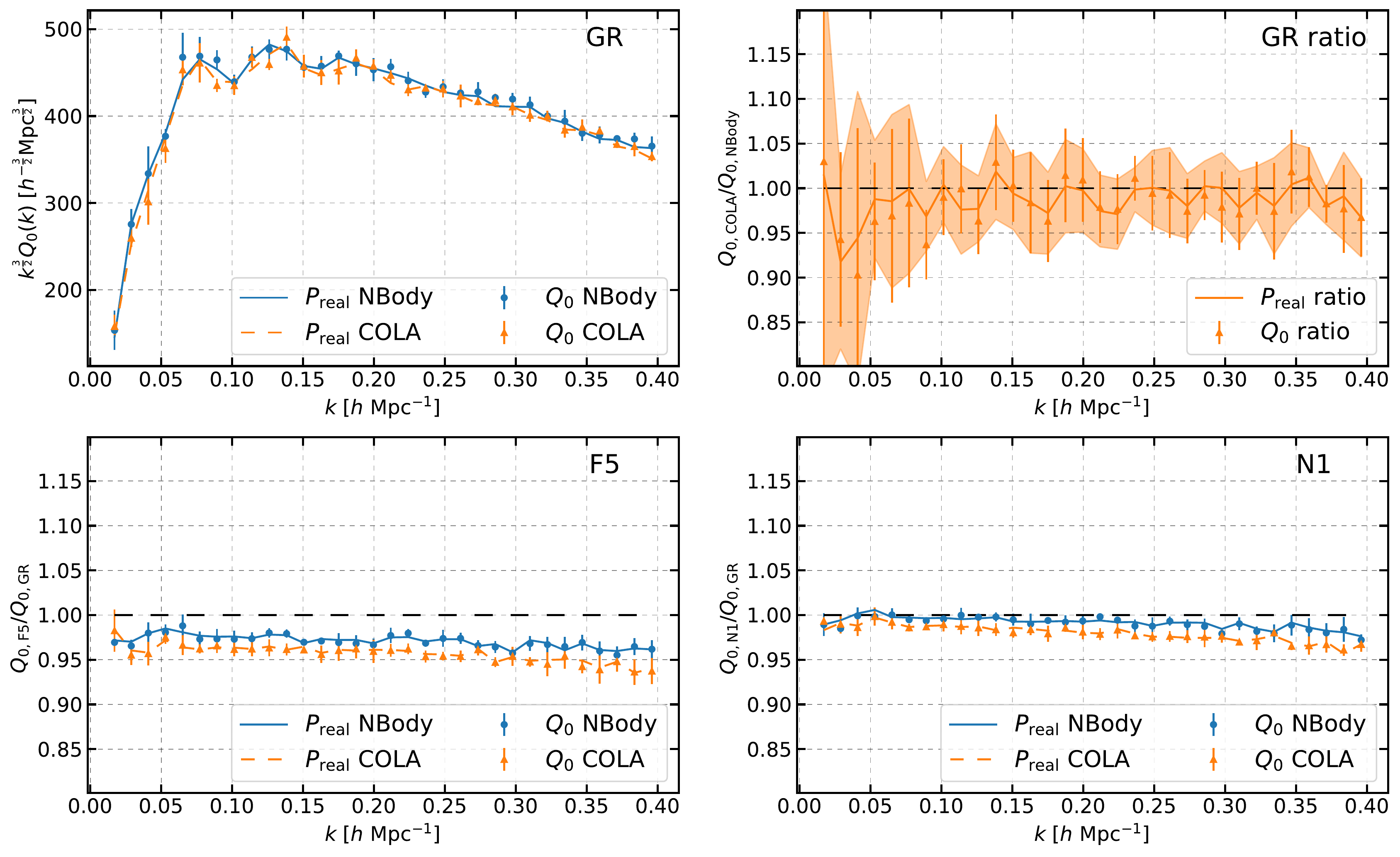}
\caption{\label{fig:Q0_halos} Comparison of the power spectrum orthogonal to the line-of-sight $Q_0$ (dots and error-bars) and the real space power spectrum $P^{r}$ (lines and shaded regions) for halos in COLA (in orange) and {\it N}-body simulations  (in blue). \textit{Top left:} Full signal in GR. \textit{Top right:} Ratio of the GR signal between COLA and {\it N}-body simulations. \textit{Bottom:} Boost factors, the ratio of $Q_0$ in F5 and N1 to that in GR.}
\end{figure}

\begin{figure}
\centering
\includegraphics[width=.98\textwidth]{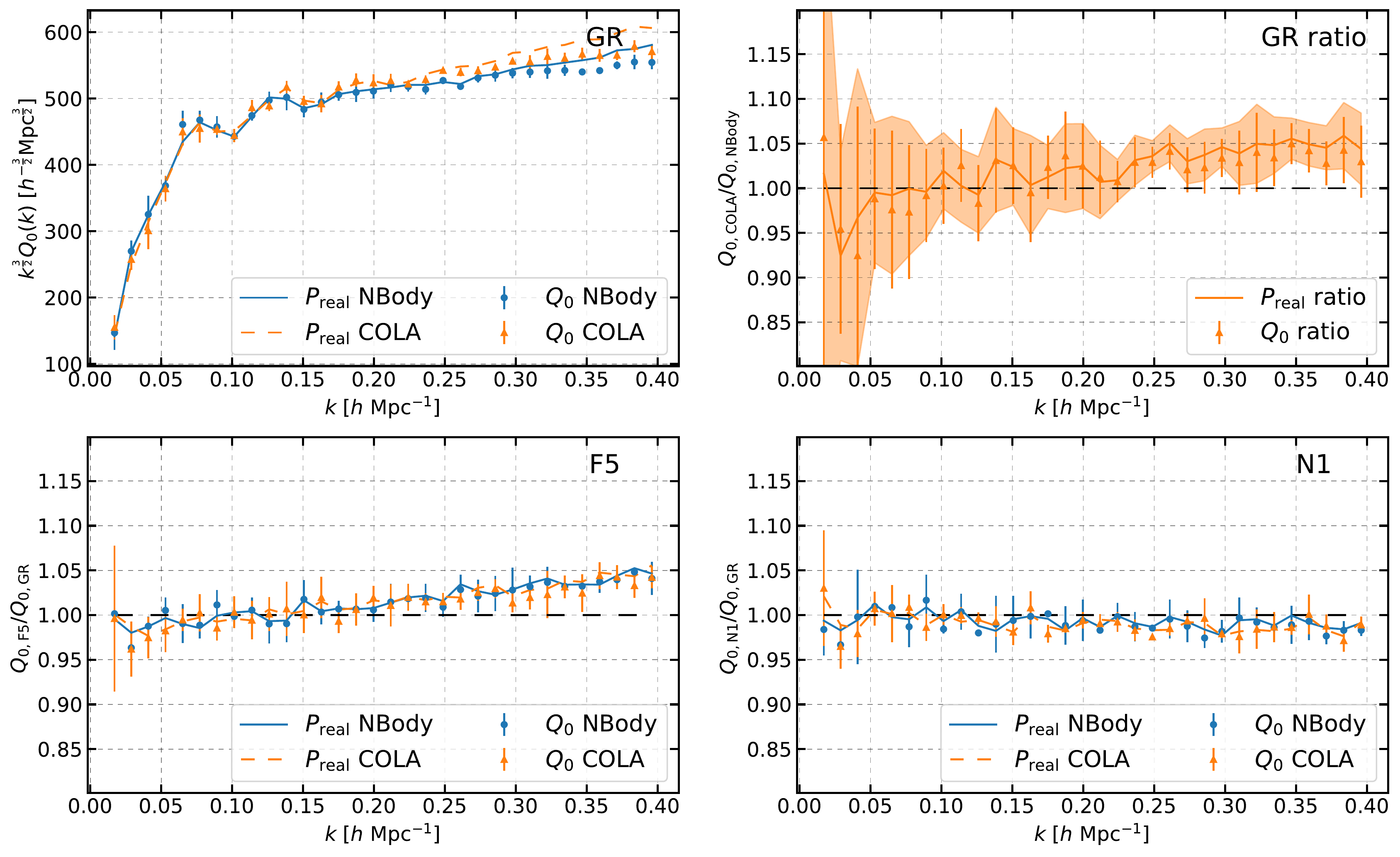}
\caption{\label{fig:Q0_galaxies}Same as figure~\ref{fig:Q0_halos} but for galaxies}
\end{figure}

We compute the real space power spectrum and the multipoles in redshift-space in our halo and galaxy catalogues with the publicly available code \codeword{nbodykit}\footnote{\url{https://nbodykit.readthedocs.io}}. We assign the density to a $512^3$ mesh grid with the Triangular Shaped Clouds (TSC) interpolation and we correct the resulting power spectra for shot noise and the effects of the interpolation used \cite{Jing:2004fq,Sefusatti:2015aex}. We use bins of width $\Delta k=2 k_f$ between $k_{\rm min}=\frac{3}{2} k_f$ and $k_{\rm max}= 0.4 \hompc$, where $k_f$ is the fundamental frequency of the box. For the multipoles in redshift-space, we add RSD to the real space positions of tracers using in turn lines-of-sight parallel to each of the simulation box axes, and average the results over the three lines-of-sight. In the case of galaxy catalogues, we also average over the 5 HOD realisations for each simulation realisation. We then use the mean and standard deviation over the simulations' realisations to produce the signals and error bars that we plot in figure~\ref{fig:Q0_halos} and figure~\ref{fig:Q0_galaxies}.

In figure~\ref{fig:Q0_halos} we compare $Q_0$ to the real space power spectrum for halos. The top left panel shows that $Q_0$ is in good agreement with $P^{r}$ both in COLA and {\it N}-body halo catalogues. The top right panel shows that COLA and {\it N}-body are consistent within the variance for both $Q_0$ and $P^{r}$. The bottom panels show that the MG boost factors for $Q_0$ and $P^r$ are consistent within the variance and that COLA is able to reproduce the {\it N}-body boost factors, the ratio of the power spectrum between MG and GR, with $\sim2\%$ accuracy at all scales in both F5 and N1. The MG signal in F5 is in general stronger than in N1 where the signal is compatible with GR within the variance.

The comparison between $Q_0$ and $P^{r}$ for the galaxy catalogues is shown in figure~\ref{fig:Q0_galaxies}. The top left panel shows that $Q_0$ is in good agreement with $P^{r}$ up to $k\sim 0.25 \hompc$ in COLA and up to $k\sim 0.3 \hompc$ in {\it N}-body galaxy catalogues, where it starts deviating due to the truncation of the sum in eq.~\eqref{Q0ofP024}. The top right panel shows that COLA and {\it N}-body are consistent within the variance for both $Q_0$ and $P^{r}$ up to $k\sim 0.24 \hompc$ where COLA starts deviating from {\it N}-body. %
The bottom panels show that the MG boost factors for $Q_0$ and $P^r$ are consistent within the variance at all scales as are COLA and {\it N}-body results. We note that, in spite of the limits of the truncation in eq.~\eqref{Q0ofP024} and of the precision of COLA in reproducing the {\it N}-body power spectrum in GR, the MG boost factors are fully consistent at all scales. Overall the MG signal is quite weak in N1 while it reaches the $5\%$ in F5 on small scales.

This suggests that the observable $Q_0$ can be an interesting statistic to include in cosmological analysis to constrain MG theories and that COLA can be used to accurately predict the MG boost factors of $Q_0$. Furthermore, the good agreement of the MG boost factors of $P^r$ and $Q_0$ estimated using only the multipoles $P_0$, $P_2$, $P_4$ suggests that higher order multipoles can be modelled using the GR theory without significantly impacting the MG signal.  
\section{Bispectrum}
\label{sec:Bispectrum}
The bispectrum is the simplest higher-order summary statistics beyond the power spectrum and it holds non-gaussian information of the density field. The study of bispectrum also serves as a test for the screening approximation adopted in COLA simulations.

\subsection{Definitions and measurements}
We define the bispectrum as the three-point function in Fourier space for closed triangle configurations
\begin{equation}
    \left\langle\delta\left(\boldsymbol{k}_{1}\right) \delta\left(\boldsymbol{k}_{2}\right) \delta\left(\boldsymbol{k}_{3}\right)\right\rangle \equiv(2 \pi)^{3} B\left(k_{1}, k_{2}, k_{3}\right) \delta_{D}\left(\boldsymbol{k}_{1}+\boldsymbol{k}_{2}+\boldsymbol{k}_{3}\right) \,.
\end{equation}
and the reduced bispectrum as the ratio
\begin{equation}
    Q(k_{1}, k_{2}, k_{3})=\frac{B(k_{1}, k_{2}, k_{3})}{P(k_{1}) P(k_{2})+P(k_{2}) P(k_{3})+P(k_{3}) P(k_{1})} \, ,
\end{equation}
where $P(k)$ is the power spectrum \cite{Scoccimarro:2000sn}. In the absence of primordial non-gaussianity, the bispectrum at the leading order in perturbation theory shows a $P^2$ scaling which is removed in the reduced bispectrum to highlight the information content beyond the power spectrum \cite{Fry:1984}.

To measure the bispectrum in our simulations we use the bispectrum estimator proposed in \cite{Scoccimarro:2015bla} as implemented in the publicly available library \codeword{FML}\footnote{\href{https://github.com/HAWinther/FML}{https://github.com/HAWinther/FML}}. The density is interpolated on a $360^3$ cells grid with a 4th order interpolation scheme known as Piece-wise Cubic Spline (PCS) with interlacing \cite{Scoccimarro:2015bla,Sefusatti:2015aex}. We estimate the bispectrum in bins of size $\Delta k = 3 k_f$, where $k_f = 2\pi/L$ is the fundamental frequency of the box, centring the first bin in $k_{\rm min} = 3 k_f$. The shot noise, which is subtracted from the bispectrum signal, is calculated as 
\begin{equation}
    B^{\mathrm{SN}}\left(k_{1}, k_{2}, k_{3}\right)=\frac{1}{\bar{n}}\left(P\left(k_{1}\right)+P\left(k_{2}\right)+P\left(k_{3}\right)\right)+\frac{1}{\bar{n}^{2}} \, ,
\end{equation}
with $\bar{n}$ being the number density of tracers and $P(k)$ their power spectrum \cite{Matarrese:1997sk}.

To get an insight into the dependence of the bispectrum signals on the triangle shape, we bin all the configurations with $k_1>0.1 \hompc$ in a grid of $x_2 \equiv k_2/k_1$, $x_3 \equiv k_3/k_1$, and average the signals weighted with the number of fundamental triangles falling in each bin
\begin{equation}\label{BispecConf}
    \overline{B}(x_2^i, x_3^j) = \frac{\sum_{k_1= k_{\rm min}}^{k_{\rm max}} \sum_{k_2,k_3}^{\rm bin} B(k_1, k_2, k_3) N^{\mathrm{T}}(k_1, k_2, k_3)}{\sum_{k_1= k_{\rm min}}^{k_{\rm max}} \sum_{k_2,k_3}^{\rm bin} N^{\mathrm{T}}(k_1, k_2, k_3)} \, ,
\end{equation}
where $N^{\mathrm{T}}$ is the number of fundamental closed-triangles
\begin{equation}
    N^{\mathrm{T}}(k_1, k_2, k_3)=\prod_{i=1}^{3} \int_{k_{i}} d^{3} q_{i} \delta_{\mathrm{D}}\left(\Vec{q_1}+\Vec{q_2}+\Vec{q_3}\right) \sum_{n_1,n_2,n_3} \delta_{\mathrm{D}}(\Vec{q_1}-n_1 k_f)\delta_{\mathrm{D}}(\Vec{q_2}-n_2 k_f) \delta_{\mathrm{D}}(\Vec{q_3}-n_3 k_f)\, ,
\end{equation}
and similarly for the reduced bispectrum
\begin{equation}\label{RedBispecConf}
    \overline{Q}(x_2^i, x_3^j) = \frac{\sum_{k_1= k_{\rm min}}^{k_{\rm max}} \sum_{k_2,k_3}^{\rm bin} Q(k_1, k_2, k_3) N^{\mathrm{T}}(k_1, k_2, k_3)}{\sum_{k_1= k_{\rm min}}^{k_{\rm max}} \sum_{k_2,k_3}^{\rm bin} N^{\mathrm{T}}(k_1, k_2, k_3)}.
\end{equation}

\subsection{Results for bispectrum}
\begin{figure}
\centering
\includegraphics[width=.98\textwidth]{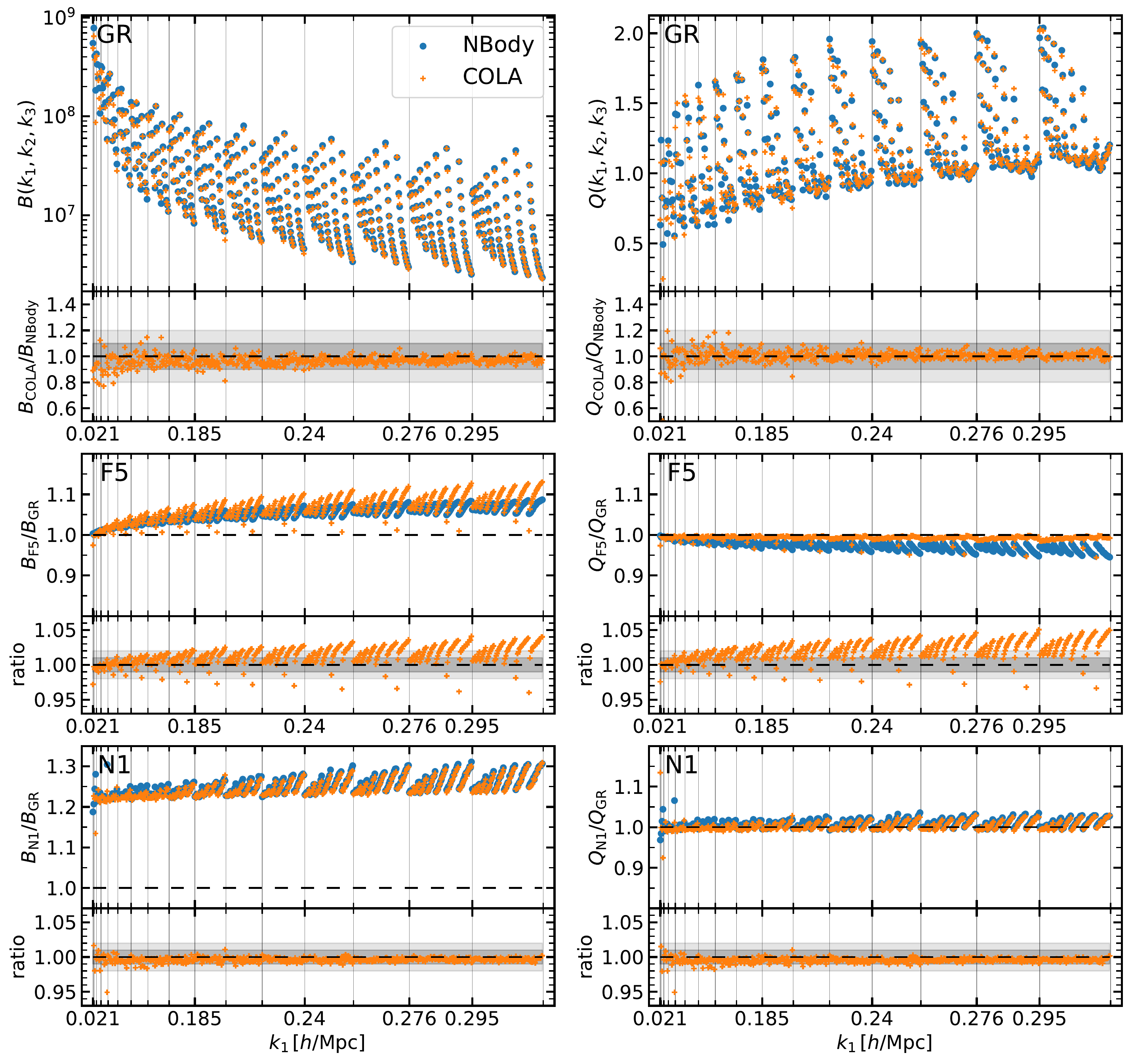}
\caption{\label{fig:DM_bispec} Comparison of bispectrum (left column) and reduced bispectrum (right column) of the matter distribution in COLA (orange crosses) and {\it N}-body (blue dots) simulations. The configurations with $k_1 \ge k_2 \ge k_3$ are displayed in ascending order of the values of the wavenumbers $k_1$, $k_2$, and $k_3$ respectively. The vertical lines denote the value of $k_1$ for the configurations immediately to the right of each line. \textit{Top:} Full signal in GR. \textit{Middle and Bottom:} Boost factors in F5 and N1 respectively. In each panel, the bottom sub-panel shows the ratio between the COLA and {\it N}-body signals displayed in the top sub-panel.}
\end{figure}

\begin{figure}
\centering
\includegraphics[width=.98\textwidth]{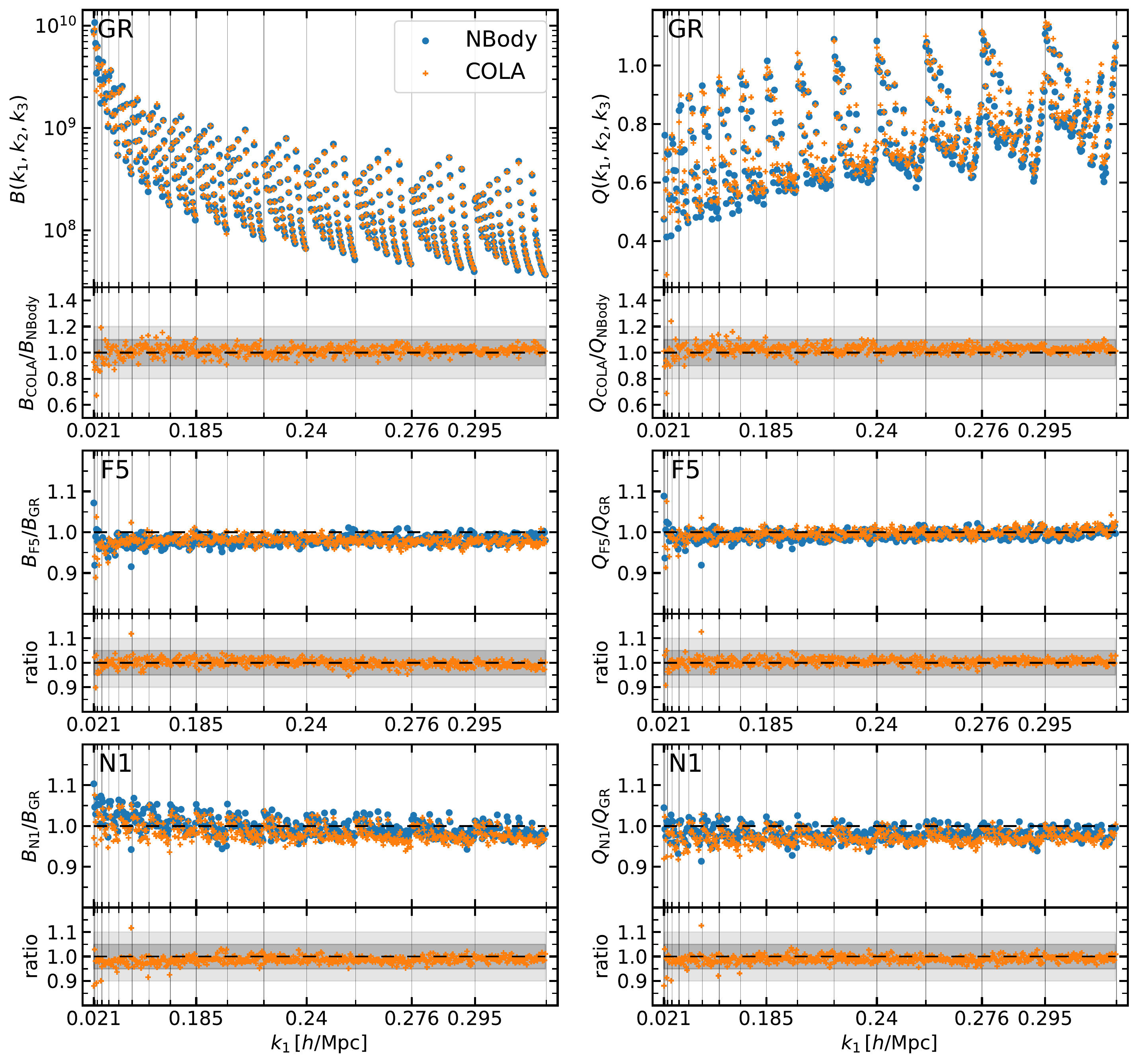}
\caption{\label{fig:Galaxy_bispec_RS} Same as figure~\ref{fig:DM_bispec} but for the monopole of bispectrum and reduced bispectrum of galaxies in redshift-space.}
\end{figure}

Since the bispectrum is a function of three variables, it is non-trivial to represent it in 2D plots, so we follow an approach often used in literature and represent all the configurations of the bispectrum in a single scatter plot by restricting our analysis to the configurations with $k_1 \ge k_2 \ge k_3$ and organising them in ascending order of the values of the wavenumbers $k_1$, $k_2$, and $k_3$, respectively. We show all the configurations of the bispectrum (and the reduced bispectrum) in figure~\ref{fig:DM_bispec} and figure~\ref{fig:Galaxy_bispec_RS} for DM and galaxies in redshift-space respectively. We show also the plots for halos and galaxies in real space in appendix~\ref{sec:AppA}.
In these figures, the left column shows the full bispectrum signal and the right column the reduced bispectrum. We split the results into three rows, the top one shows the full signal in GR, the middle one shows the boost factor in F5 (i.e. the ratio between F5 and GR) and the bottom one shows the boost factor in N1. To benchmark results with respect to full {\it N}-body, each row contains a sub-panel at the bottom showing the ratio between the COLA and {\it N}-body signals.

The MG boost factors of $\bar{B}$ and $\bar{Q}$ (from eq.~\eqref{BispecConf} and eq.~\eqref{RedBispecConf} respectively), 
that is, the ratio of $\bar{B}$ and $\bar{Q}$ between MG and GR, are used to produce figure~\ref{fig:DM_bispec_conf} for DM and figure~\ref{fig:Galaxy_bispec_RS_conf} for galaxies in redshift-space (see appendix~\ref{sec:AppA} for equivalent plots of halos and galaxies in real space). In each figure, the left panels display the full bispectrum and the right panels the reduced bispectrum. We show the results for  
F5 in the top row and the results for N1 in the bottom row. The colour represents the amplitude of the MG signal, according to the colour bar. 

\begin{figure}
\centering
    \subfloat[][Full Bispectrum]{
    \includegraphics[width=.48\textwidth,clip]{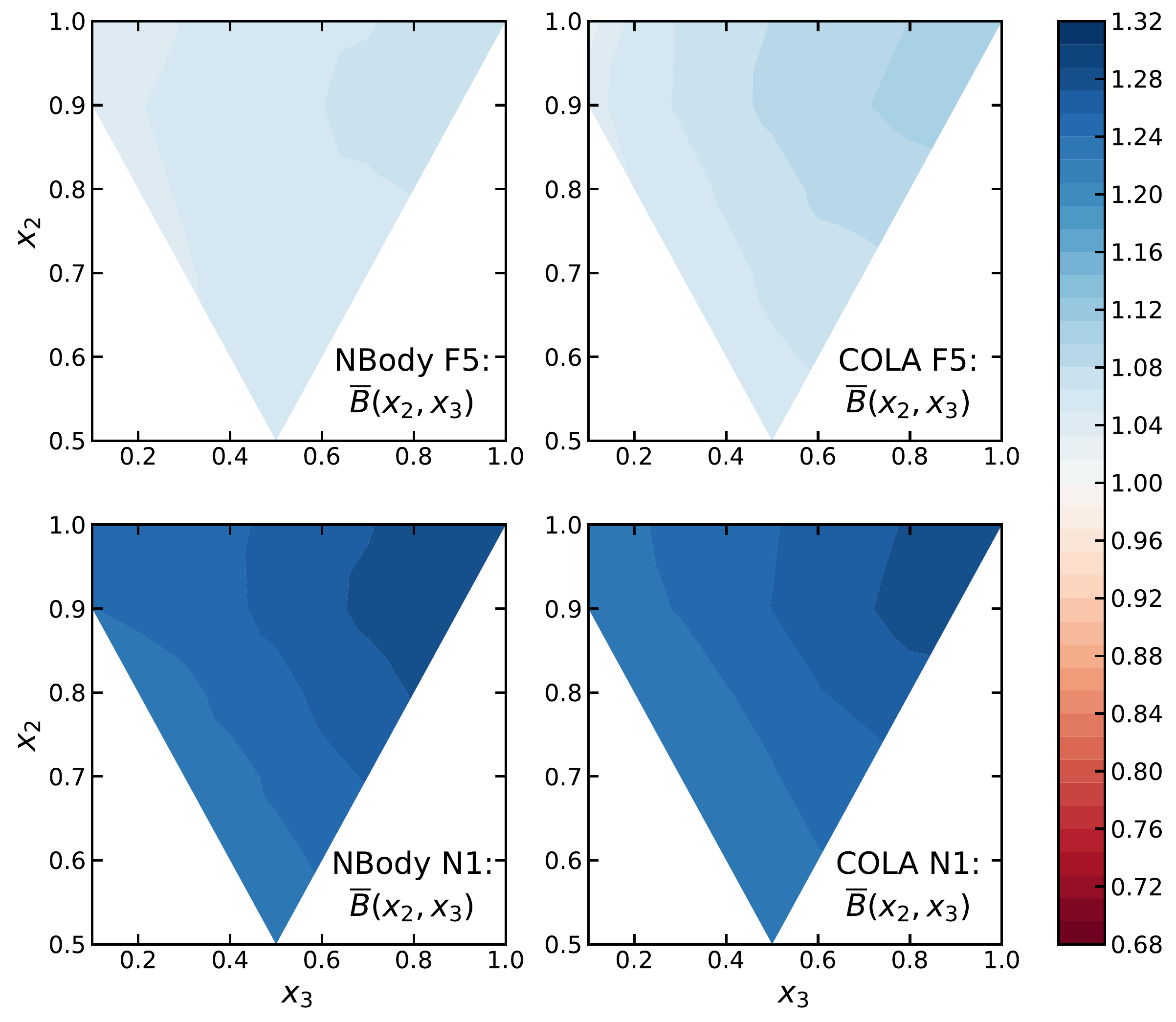}
    }
    \hfill
    \subfloat[][Reduced Bispectrum]{
    \includegraphics[width=.48\textwidth]{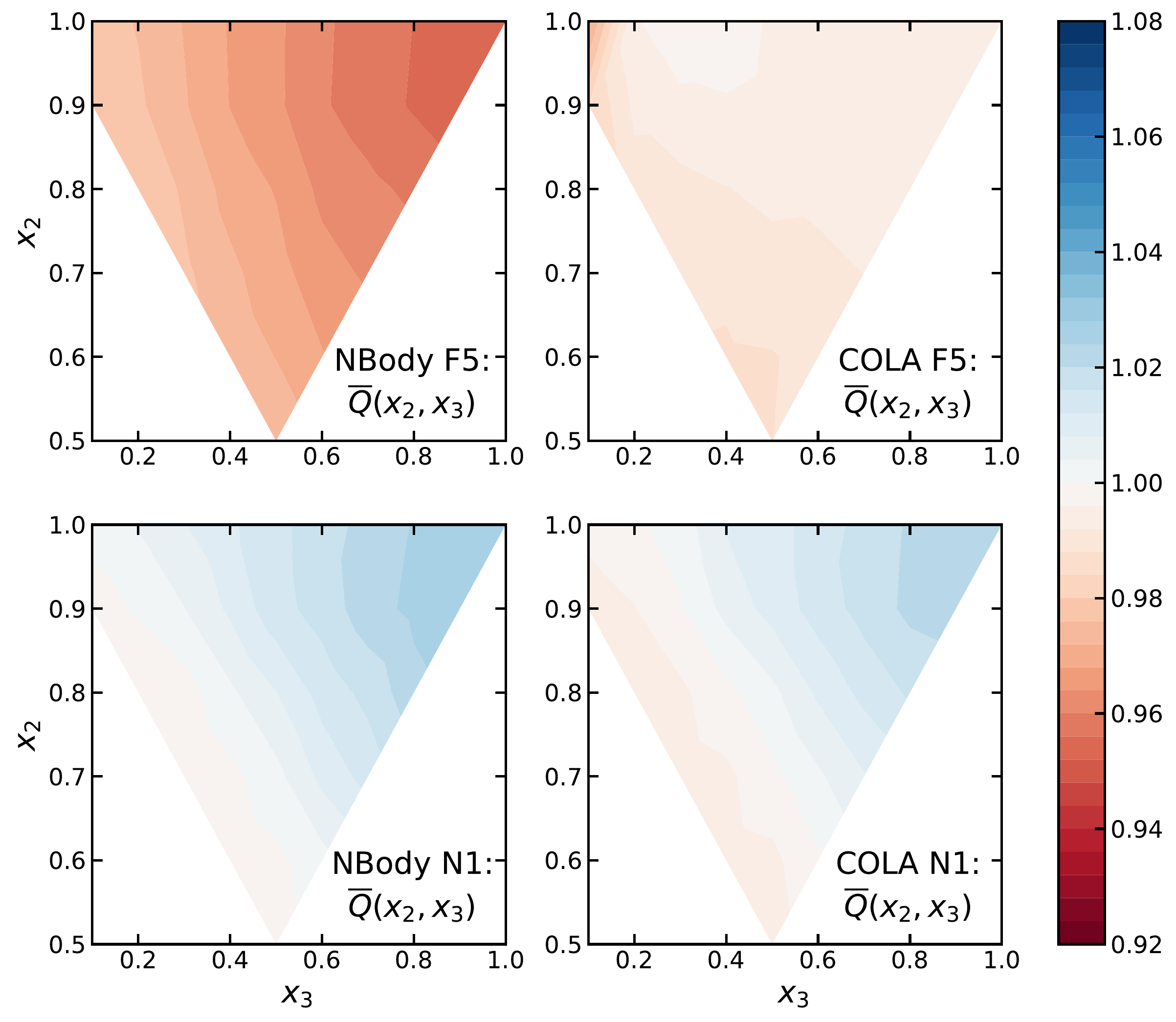}
    }
\caption{\label{fig:DM_bispec_conf}Comparison of the configuration dependence of the F5 (top row) and N1 (bottom row) boost factors of bispectrum (on the left) and reduced-bispectrum (on the right) of DM in {\it N}-body (first and third columns) and COLA simulations (second and fourth columns). The colour bars show the amplitude of the boost factors with blue (red) denoting stronger (weaker) signal in MG with respect to GR. \textit{In each panel:} The top right, top left and bottom corners of the triangle correspond to the equilateral, squeezed and folded configurations respectively. The squeezed configuration is missing from the figure due to the bin's width.}
\end{figure}

\begin{figure}
\centering
    \subfloat[][Full Bispectrum]{
    \includegraphics[width=.48\textwidth,clip]{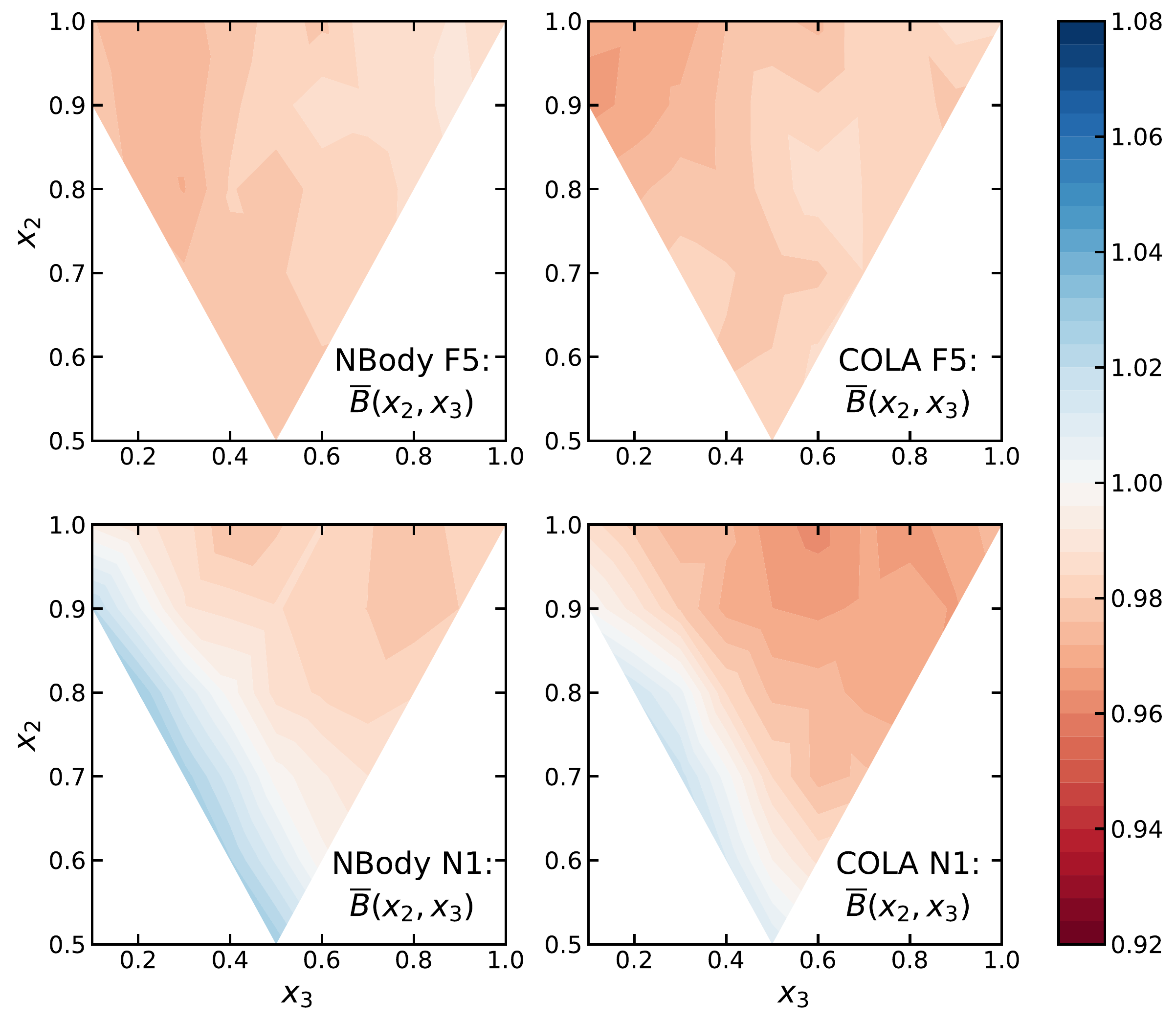}
    }
    \hfill
    \subfloat[][Reduced Bispectrum]{
    \includegraphics[width=.48\textwidth]{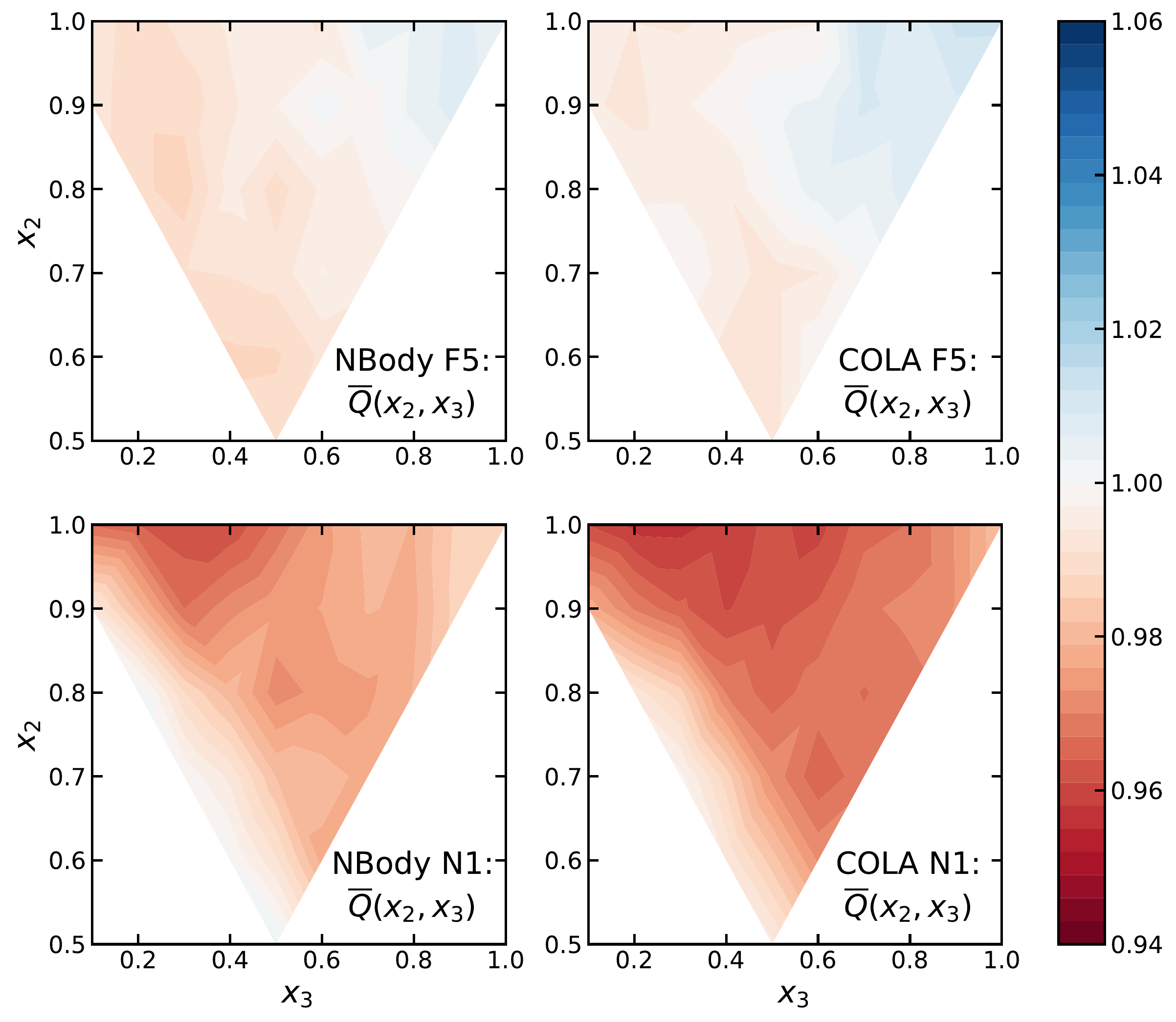}
    }
\caption{\label{fig:Galaxy_bispec_RS_conf}Same as figure~\ref{fig:DM_bispec_conf} but for the monopole of the bispectrum and reduced bispectrum of galaxies in redshift-space.}
\end{figure}

At the DM level, figure~\ref{fig:DM_bispec} shows that COLA is able to reproduce {\it N}-body results in GR with $\sim 10\%$ accuracy both for the bispectrum and the reduced bispectrum in agreement with the findings in \cite{Colavincenzo_2018}. Focusing on the MG boost factors in {\it N}-body, we notice that the strong MG signal of the bispectrum, reaching $\sim 30\%$ in N1 and $\sim 10\%$ in F5, is suppressed in the reduced bispectrum. This shows that most of the bispectrum signal is due to the enhancement of the power spectrum in MG. The MG boost factors of COLA do not reproduce the information content beyond the power spectrum as highlighted by the reduced bispectrum signal in the case of F5, while it reproduces the {\it N}-body signals with $\sim 2\%$ accuracy in N1. The failure of COLA simulations in F5 gravity can be traced back to the use of the screening approximation, which allows us to recover with good accuracy the power spectrum but produces a DM bispectrum signal consistent with the results of \cite{Gil-Marin:2011iuh} in the case of simulations without the chameleon mechanisms (see their figure 1). 
This is because the screening approximation used in COLA simulations linearises the scalar field equation and thus it lacks the contribution to the bispectrum from the non-linearity of the scalar field. This effect is stronger in F5, where the effect of screening is strong even on quasi-non-linear scales, while it is weak in N1. Concerning the configuration dependence of the DM bispectrum, figure~\ref{fig:DM_bispec_conf} shows that the bispectrum boost factor signals of DM are stronger for equilateral configurations. COLA results reproduce the same configuration dependence of {\it N}-body in N1, but the reduced bispectrum boost factor in F5 clearly shows the failure of COLA simulations in reproducing non-trivial 3-points interactions, due to the screening approximation used in COLA.

Figure~\ref{fig:Galaxy_bispec_RS} shows the bispectrum monopole of the galaxy catalogues in redshift-space. The agreement between COLA and {\it N}-body is within $10\%$ in GR for both the bispectrum and the reduced bispectrum. In the F5 and N1 gravity models, the bispectrum boost factors show deviations from GR at $\sim5\%$ level for some configurations, and the same is true for the reduced bispectrum. COLA reproduces {\it N}-body results for the boost factors with at least $5\%$ accuracy. 
The much weaker MG signal of the redshift-space galaxy bispectrum monopole compared to the DM bispectrum is attributable to the HOD parameters tuning procedure, which tries to reproduce the target multipoles of the galaxy power spectrum, in this case the GR one \cite{Fiorini:2021dzs}. This tuning produces a power spectrum contribution to the bispectrum that is similar for GR and MG theories (as can be seen from the boost-factors of the bispectrum in the left column) and, on the other hand, dilutes the MG signal of the bispectrum coming from higher-order contributions of DM clustering due to the nonlinearity of galaxy bias (as evidenced by the boost-factors of the reduced bispectrum in the right column). This influence of galaxy bias is also visible in figure~\ref{fig:Galaxy_bispec_RS_conf} that shows the configuration dependence of galaxy bispectra boost factors substantially differs from what is found in the DM bispectra boost factors. This explains why the configuration dependence in COLA is qualitatively in agreement with the {\it N}-body result. The contribution from the non-linear galaxy bias dominates over the DM non-linearity hiding the inaccuracy of COLA simulations in reproducing the DM bispectra in F5.

The bispectra of halos and galaxies in real space are affected by biases in a similar fashion to the monopole of the galaxy bispectrum in redshift space. Likewise, they do not show strong MG signals. For an equivalent analysis on the bispectra of halos and galaxies in real space see appendix~\ref{sec:AppA}.

Our results show that the MG bispectrum signal is mainly due to MG effects on the power spectrum and less to the higher order terms carrying specific information of the gravity model as shown in \cite{Gil-Marin:2011iuh} for $f(R)$ gravity. Extra care needs to be taken when using COLA simulations with screening approximations to compute the reduced bispectrum of dark matter, in particular in $f(R)$ gravity models as the non-linearity of the scalar field due to screening is neglected. This is relevant for the computation of higher-order statistics in weak lensing, for example. On the other hand, in galaxy clustering, there are large contributions from non-linear galaxy biases to the bispectrum, and this inaccuracy in the reduced bispectrum in COLA is hidden. The bispectrum, when combined with the power spectrum, is useful to break degeneracies in the galaxy biases and COLA simulations can be used to model accurately galaxy bispectrum in MG models, despite the use of screening approximations.

\section{Voids}
\label{sec:Voids}
Voids are low-density regions of space that are useful for cosmological analysis because the weaker gravitational field produces a phase-space distribution that is better described by linear models \cite{Cai:2016jek,Nadathur:2017jos}. Also, screening mechanisms do not operate efficiently in voids, thus voids can be used to test models of gravity
\cite{voidmg1,voidmg2,voidmg3,voidmg4,Cautun:2017tkc,Paillas:2018wxs,Perico:2019obq,Contarini:2020fdu}.
Observationally we cannot directly probe DM, so we have to rely on galaxies to identify voids, keeping in mind that galaxies are biased tracers of the underlying density field and that we only know the redshift-space position of galaxies.

Many void definitions have been proposed in the literature, differing in the properties (e.g. abundances and profiles) of the voids they identify. Some void definitions can reveal to be more suited than others in highlighting specific features of the underlying galaxy and DM distributions as discussed in many works (see for example \cite{Colberg:2008qg,Nadathur:2017jos,Cautun:2017tkc,Paillas:2018wxs,Massara:2022lng}).
In particular, the watershed void finding technique (implemented in various codes such as \codeword{ZOBOV} \cite{Neyrinck:2007gy} and \codeword{VIDE} \cite{Sutter:2014haa}) has been shown to identify (real-space) voids whose (stacked) galaxy density profiles in redshift-space can be approximated by linear models sensitive to the linear growth rate of structure \cite{Cai:2016jek,Nadathur:2017jos}. As MG theories may affect the linear growth rate $f$, we want to test if $f$ can also be recovered in MG theories using the RSD model for the void-galaxy Cross-Correlation Function (CCF) in \cite{Nadathur:2017jos}.

In subsection \ref{ssec:RSDinVoids} and \ref{ssec:VelModel} we review the RSD model from \cite{Nadathur:2017jos,Woodfinden:2022bhx}, and in subsection \ref{ssec:VoidMeas} we perform void finding on our simulation suites and measure the void quantities relevant to the RSD model. Finally, in subsection \ref{ssec:VoidResults} we fit the multipoles of the void-galaxy CCF in redshift-space using the RSD model to recover the linear growth rate $f$. Comparing the measurements and results for the fit between COLA and {\it N}-body simulations, we validate COLA simulations for void analysis.

\subsection{Redshift-space distortion model in voids}
\label{ssec:RSDinVoids}

Let $\xi^r(\vec{r})$ be the real-space CCF between voids and galaxies. Assuming a bijective differentiable mapping between real and redshift-space $\vec{s} = \vec{f}(\vec{r})$, the redshift-space CCF $\xi^s(\vec{s})$ is related to the real-space CCF by
\begin{equation}
    \int_V \left[1+\xi^{s}(\mathbf{s})\right] d^{3} s= \int_V \left[1+\xi^{r}(\mathbf{r})\right] d^{3} r .
\end{equation}
Performing the change of variable $\vec{r} \rightarrow \vec{s}$ on the right-hand side we obtain
\begin{equation}
    \int_V \left[1+\xi^{s}(\mathbf{s})\right] d^{3} s= \int_V \left[1+\xi^{r}(\mathbf{r}(\vec{s}))\right] \mathrm{J}_{\vec{r},\vec{s}} \, d^{3} s \, ,
\end{equation}
where ${\rm J}_{\vec{r},\vec{s}}$ is the determinant of the Jacobian matrix $\frac{\partial \vec{r}}{\partial \vec{s}}$. Requiring that this equation holds for every arbitrary volume $V$ implies
\begin{equation}
        \left[1+\xi^{s}(\mathbf{s})\right] =  \left[1+\xi^{r}(\mathbf{r}(\vec{s}))\right] \mathrm{J}_{\vec{r},\vec{s}} \, .
\end{equation}
This relation can be manipulated to express the redshift-space CCF $\xi^s$ in terms of $\xi^r$, once the mapping between real and redshift-space is fixed, determining the two unknowns $\mathbf{r}(\vec{s})$ and $\mathrm{J}_{\vec{r},\vec{s}}$.
This mapping in the distant observer approximation can be described by
\begin{equation} \label{RSD_map}
    \mathbf{s}=\mathbf{r}+\frac{\mathbf{v} \cdot \hat{\mathbf{z}}}{a H} \hat{\mathbf{z}},
\end{equation}
where $\vec{v}$ is the galaxy velocity, $a$ is the scale factor, and $H$ is the Hubble parameter. The void positions are by construction assumed to be invariant under redshift-space remapping, since the CCF in this model always refers to the correlation between real-space voids and galaxies in either real or redshift-space \cite{Nadathur:2017jos}.

Assuming a spherically symmetric model for the galaxy velocity $\vec{v} = v(r)\vec{\hat{r}}$, the Jacobian determinant is given by
\begin{equation} \label{cart_J}
    {\rm J}_{\vec{r},\vec{s}} = \left[ 1+\frac{v(r)}{a H r}+\frac{z^{2} \frac{d}{d r} v\left(r\right)}{a H r^{2}}-\frac{z^{2} v(r)}{a H r^{3}} \right ]^{-1}\, ,
\end{equation} 
with $r = \norm{\vec{r}} =\sqrt{x^2 + y^2 + z^2}$, which can be rewritten as
\begin{equation}
   {\rm J}_{\vec{r},\vec{s}} = \left[1+\frac{\vbar(r)}{r}+\mu_r^{2} \frac{d \vbar(r)}{d r} -\frac{\mu_r^2 \vbar(r)}{r}\right]^{-1}  \, .
\end{equation}
where we defined $\Bar{v} \equiv \frac{v}{aH}$ and $\mu_r \equiv \frac{z}{r}$ for simplicity of notation.

In addition to the spherically symmetric model for the galaxy velocity we consider a random velocity component along the line-of-sight%
\footnote{A more realistic model would be considering a random velocity component with random orientation (and dispersion 3 times larger), but the two models are equivalent for what concerns RSD.}, %
$\vec{v} = v_r(r)\vec{\hat{r}} + v_{\parallel} \vec{\hat{z}}$, where $P(v_{\parallel}) = \mathcal{N}(0,\sigma(r))$ as proposed in \cite{Nadathur:2017jos}.
This makes the mapping between real and redshift-space depends also on $v_{\parallel}$
\begin{equation}\label{RSDmap_disp}
    \begin{aligned}
        \mathbf{s}(\vec{r}, v_{\parallel})&=\mathbf{r}+\frac{\left(v_r(r)\vec{\hat{r}} + v_{\parallel} \vec{\hat{z}}\right) \cdot \hat{\mathbf{z}}}{a H} \, \hat{\mathbf{z}} \\
        &=\mathbf{r}+\frac{\left(v_r(r)\mu_r + v_{\parallel} \right)}{a H} \, \hat{\mathbf{z}}\, ,
    \end{aligned}
\end{equation}
but does not explicitly affect the Jacobian, as $v_{\parallel}$ is an integration variable independent of the real-space radius $r$.
We solve eq.~\eqref{RSDmap_disp} for $\vec{r}$ iteratively to obtain $\vec{r}(\vec{s}, v_{\parallel})$ and integrate over $v_{\parallel}$ to compute the theoretical prediction for the redshift-space CCF
\begin{equation}
    \xi^{s}_{\rm th}(s, \mu_s) =  \int\left(1+\xi^{r}\right) \mathrm{J}_{\vec{r},\vec{s}} e^{-\frac{v_\parallel^2}{2 \sigma_v^2}}\, dv_{\parallel} -1 \, ,
\end{equation}
where $\xi^r$, $\mathrm{J}_{\vec{r},\vec{s}}$ and $\sigma_v$ are expressed in terms of $s$ and $\mu_s$.
Finally we project $\xi^{s}_{\rm th}(s, \mu_s)$ on the Legendre polynomials to compute monopole and quadrupole of the void-galaxy CCF in redshift-space
\begin{equation} \label{theory_xiell}
\begin{gathered}
    \xi^{s}_{0,{\rm th}}(s) = \int_{0}^{1} \xi^{s}_{\rm th}(s, \mu_s) \, d\mu_s \, ,\\ 
    \xi^{s}_{2,{\rm th}}(s) = \frac{5}{2}\int_{0}^{1} \xi^{s}_{\rm th}(s, \mu_s) \left(3 \mu_s^{2}-1\right)\, d\mu_s \,.
\end{gathered}
\end{equation}

In the comparisons of the void-galaxy CCF in redshift-space shown in subsection~\ref{ssec:VoidMeas}, we include these theoretical predictions without any further assumption to test the accuracy of this approach.

\subsection{Velocity model}
\label{ssec:VelModel}
Galaxy velocities correlate with void positions, as galaxies are attracted towards overdensities far from the void centre. In an isotropic universe, despite that voids have in general very irregular shapes, the average matter distribution of stacked voids is isotropic and therefore spherically symmetric. The average velocity of galaxies around voids depends on the matter distribution and it is also spherically symmetric.
This velocity can be estimated in simulations and it has been found to be well described down to scales of $\sim 30$ Mpc$/h$ by the simple linear model
\begin{equation}\label{VelModel}
    v_{r}^{\text {model }}(r)=-\frac{1}{3} f a H r \Delta(r) \, ,
\end{equation}
where $f$ is the linear growth rate and 
$
    \Delta(r) \equiv \frac{3}{r^{3}} \int_{0}^{r} \delta(r') r^{\prime 2} d r'
$
is the integrated matter density profile \cite{Peebles:1994xt,Nadathur:2017jos}. We will show in subsection~\ref{ssec:VoidMeas} a comparison between this velocity model and the radial-velocity profile estimated from our simulations that confirms the same trend of \cite{Nadathur:2017jos} for the accuracy of the velocity model.

\subsection{Void finding and measurements}\label{ssec:VoidMeas}
\begin{figure}
\centering
\includegraphics[width=.98\textwidth]{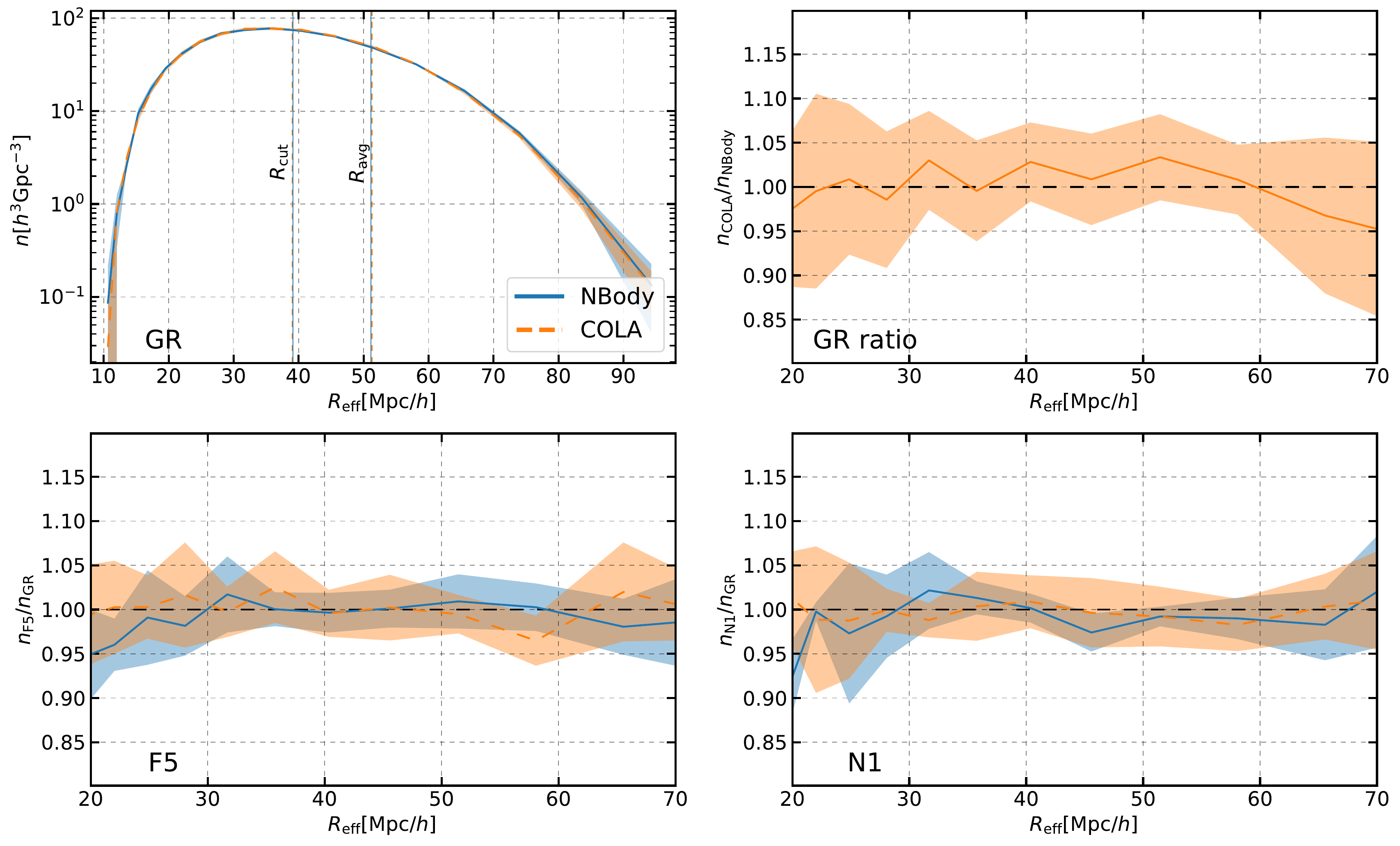}
\caption{\label{fig:VoidSizeFun}Comparison of the void size function in COLA (in orange) and {\it N}-body simulations (in blue). \textit{Top left:} Full signal in GR. The vertical lines denote the minimum and average void radii after applying the median cut. \textit{Top right:} Ratio of the GR signal between COLA and {\it N}-body simulations. \textit{Bottom:} Boost factors in F5 and N1. The panels involving ratios are shown in the range $[20,70] \mpcoh$ as the signal-to-noise ratio deteriorates outside of this range.}
\end{figure}

To perform void-finding for the real-space galaxy mocks, we make use of the publicly available code \codeword{Revolver}\footnote{\href{https://github.com/seshnadathur/Revolver}{https://github.com/seshnadathur/Revolver}} which implements two void-finding techniques: \codeword{ZOBOV} \cite{Neyrinck:2007gy} and \codeword{voxel}. With the aim of performing a similar analysis to \cite{Nadathur:2017jos} we choose \codeword{ZOBOV} void-finding. This algorithm finds voids as the watershed basins of the density field of galaxies in real-space estimated using the Voronoi tessellation \cite{Neyrinck:2007gy}. The void effective radius is determined by the void total volume $V$ via the formula $R_{\rm eff}=(3 V / 4 \pi)^{1 / 3}$. We use this definition of void radius to count the voids in 20 bins logarithmically spaced between 10 and 100 $\mpcoh$ to obtain the void size function displayed in figure~\ref{fig:VoidSizeFun}. This shows that the void-size functions in GR, N1 and F5 are all compatible within the variance, as are the void-size functions of COLA and {\it N}-body simulations. 

We select a sub-sample of voids whose radius is larger than the median radius of the catalogue, which is sufficient to select voids that are uncorrelated and statistically isotropic \cite{Nadathur:2017jos}. %
This results in void catalogues with a minimum void radius $R_{\rm cut} \simeq 39 \mpcoh$ and an average void radius $R_{\rm avg} \simeq 51 \mpcoh$ across all the simulations. We add the vertical lines in the top-left panel of figure~\ref{fig:VoidSizeFun} to highlight these two scales in the void-size function. The void centre is identified as the centre of the largest empty sphere inside the void volume. This is done by finding the circumcentre of the tetrahedron described by the 4 galaxies closest to the under-density peak.
Given the definition of the void centre, it is possible to study several quantities characterising the profile of voids. In particular, we focus on: 
\begin{itemize}
    \item the integrated matter density profile, $\Delta(r) \equiv \frac{3}{r^{3}} \int_{0}^{r} \delta(r') r^{\prime 2} d r'$, where $\delta(r)$ is the matter density-contrast profile of stacked voids;
    \item the galaxy radial-velocity profile, $v_{\rm r}(r) = \left\langle \vec{v}_g \cdot \hat{r} \right\rangle$;
    \item the galaxy velocity-dispersion profile $\sigma_{v_{\parallel}}(r) = \nicefrac{\sqrt{ \left\langle |\vec{v}_g |^2\right\rangle - v_{\rm r}^2 }}{ \sqrt{3}}$;
    \item the CCF of voids and galaxies in real and redshift-space, $\xi^r$ and $\xi^s$ respectively.
\end{itemize} 
These quantities, relevant to the RSD model of the void-galaxy CCF discussed in subsection~\ref{ssec:RSDinVoids}, are estimated using 25 linearly spaced radial bins from the void centre to $120 \mpcoh$. To measure the integrated-density profile and the CCF we use the publicly available code \codeword{pyCUTE}\footnote{\href{https://github.com/seshnadathur/pyCUTE}{https://github.com/seshnadathur/pyCUTE}} based on \codeword{CUTE}\footnote{\href{https://github.com/damonge/CUTE}{https://github.com/damonge/CUTE}} \cite{CUTE}. In the case of real-space (redshift-space) statistics, the signal is first averaged over the 5 HOD realisations (and over the three lines of sight), and then the mean and standard deviations over the 5 boxes are used respectively for the lines and shaded regions in the figures.

\begin{figure}
\centering
\includegraphics[width=.98\textwidth]{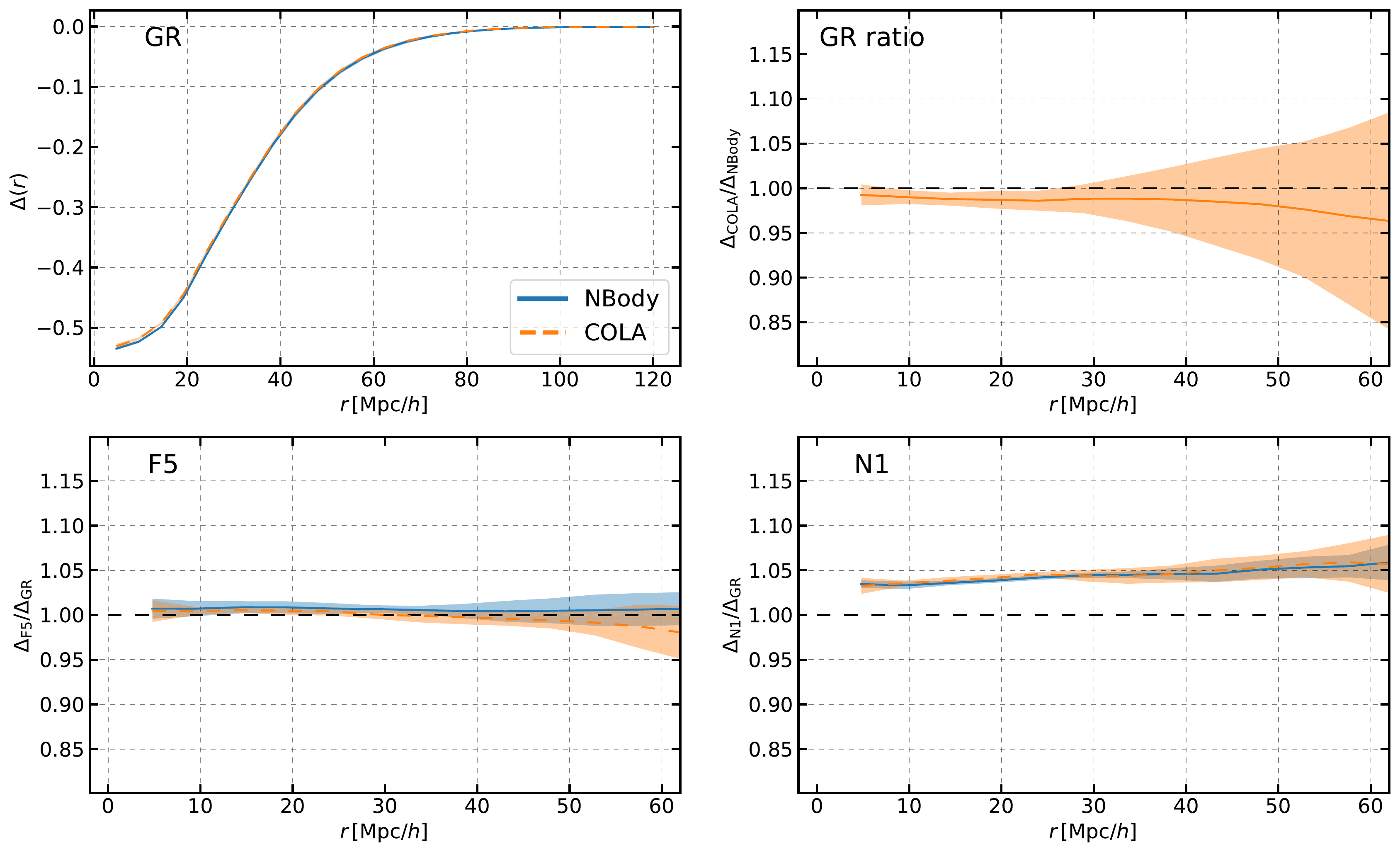}
\caption{\label{fig:DeltaProfile}Comparison of the integrated matter density profile of voids in COLA (in orange) and {\it N}-body simulations (in blue). \textit{Top left:} Full signal in GR. \textit{Top right:} Ratio of the GR signal between COLA and {\it N}-body simulations. \textit{Bottom:} Boost factors in F5 and N1. The panels involving ratios are shown up to $r\simeq 60 \mpcoh$ as the signal-to-noise ratio deteriorates for larger separations.}
\end{figure}

A comparison of the integrated matter density profile measured in our simulations is shown in figure~\ref{fig:DeltaProfile}. The signal is characterised by a good signal-to-noise ratio in the void interior (due to the amplitude of the signal and the abundance of DM tracers) which deteriorates above $\sim60 \mpcoh$ due to the signal approaching the large-scales limit of 0. COLA and {\it N}-body results are very consistent (within the variance) in all gravity models. The MG signal in the integrated-density profile is weak in the case of F5 (compatible with GR) while it shows a $\sim5\%$ enhancement in the case of N1 gravity. 

\begin{figure}
\centering
\includegraphics[width=.98\textwidth]{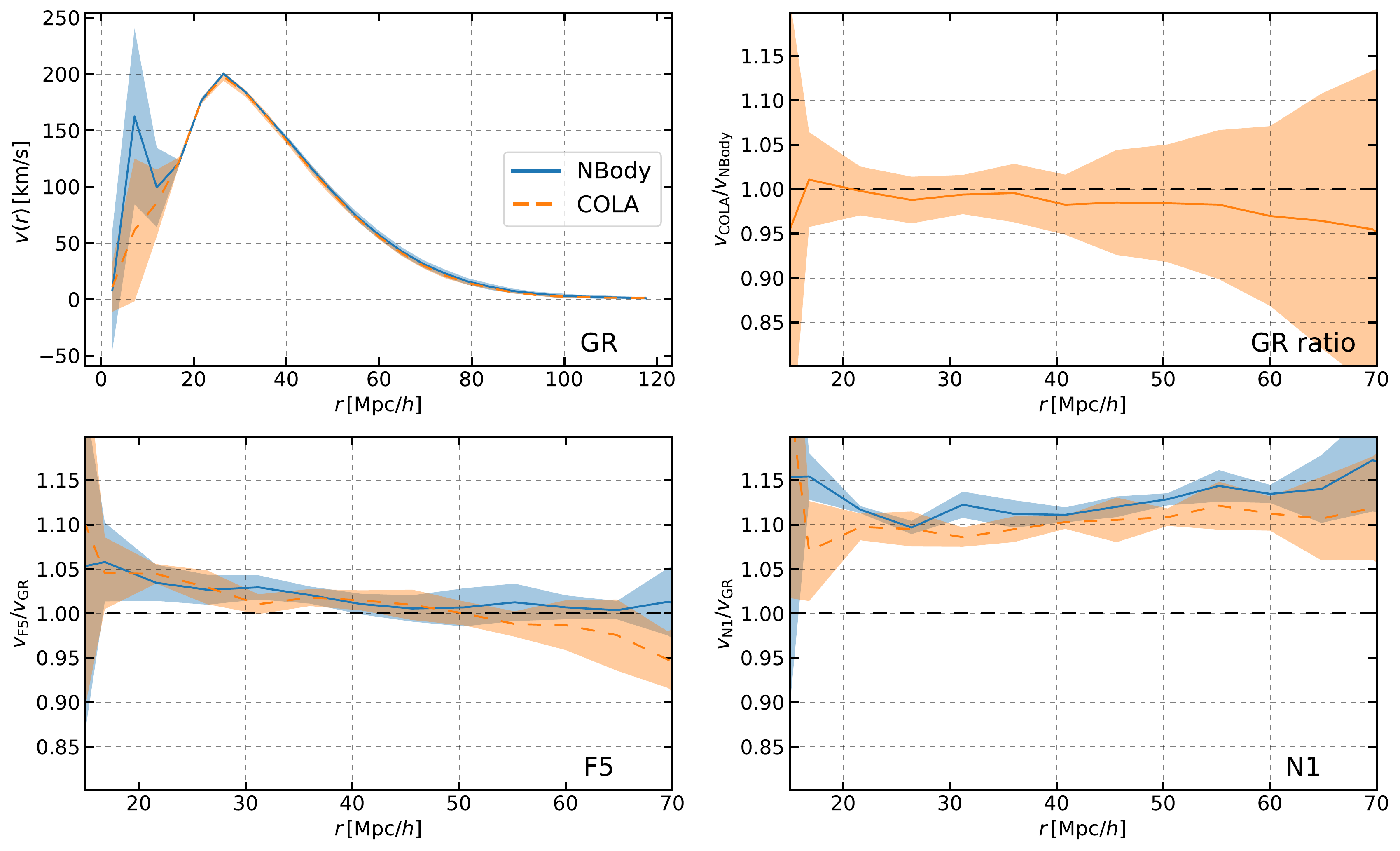}
\caption{\label{fig:VelProfile} Comparison of the galaxy radial-velocity profile around voids in COLA (in orange) and {\it N}-body simulations (in blue). \textit{Top left:} Full signal in GR. \textit{Top right:} Ratio of the GR signal between COLA and {\it N}-body simulations. \textit{Bottom:} Boost factors in F5 and N1. The panels involving ratios are displayed in the range $[20,70] \mpcoh$ as the signal-to-noise ratio deteriorates outside of the range.}
\end{figure}

The average radial velocity of galaxies as a function of the separation between the galaxies and the void centre is represented in figure~\ref{fig:VelProfile}. The bins below $15 \mpcoh$ are heavily affected by shot noise due to the low galaxy counts. %
On large scales, the signal approaching 0 is responsible for the large scatter in the ratios plots, which are therefore shown in the range $[15,70] \mpcoh$.  In GR, COLA is in agreement with {\it N}-body within the variance at all scales. A small scale-dependent MG signal is present in F5, with velocities larger than in GR for small separations and approaching GR results on large scales. The boost factor in N1 shows a $\sim10\%$ enhancement of the radial-velocity profile with respect to GR. The different behaviours of the velocity profiles in N1 and F5 are consistent with the different nature of their fifth force, long-ranged in N1, and short-ranged in F5.

\begin{figure}
\centering
\includegraphics[width=.98\textwidth]{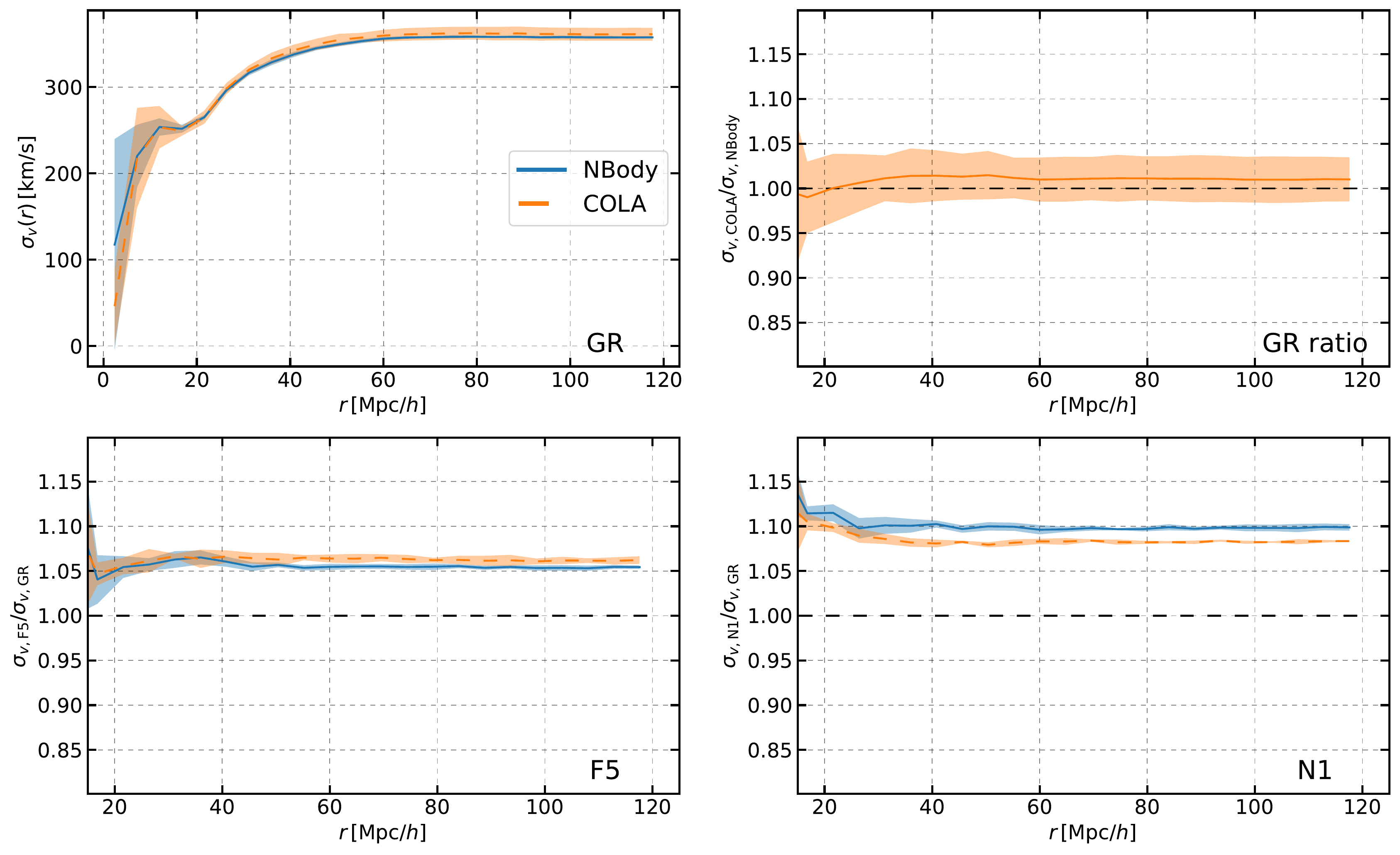}
\caption{\label{fig:VelDispProfile} Comparison of the galaxy velocity-dispersion profile around voids in COLA (in orange) and {\it N}-body simulations (in blue). \textit{Top left:} Full signal in GR. \textit{Top right:} Ratio of the GR signal between COLA and {\it N}-body simulations. \textit{Bottom:} Boost factors in F5 and N1. The panels involving ratios are displayed down to $15 \mpcoh$ as the signal-to-noise ratio deteriorates for smaller separations.}
\end{figure}

Figure~\ref{fig:VelDispProfile} displays the dispersion of the random galaxy velocity component along the line of sight. The large scatter in the first bins is again due to the low number of galaxies in the proximity of the void centre. The velocity-dispersion shows a mild dependence on separations at small separations and plateaus at $\sim 370 \, {\rm km/s}$. COLA results are in agreement with the {\it N}-body results in GR within the variance at all scales. The boost factors in F5 and N1 show a clear enhancement with respect to GR, of $\sim 5\%$ and $\sim 10\%$ respectively. The accuracy of COLA in reproducing the boost factors is $1\%$ in F5 and $2\%$ in N1.

\begin{figure}
\centering
\includegraphics[width=.98\textwidth]{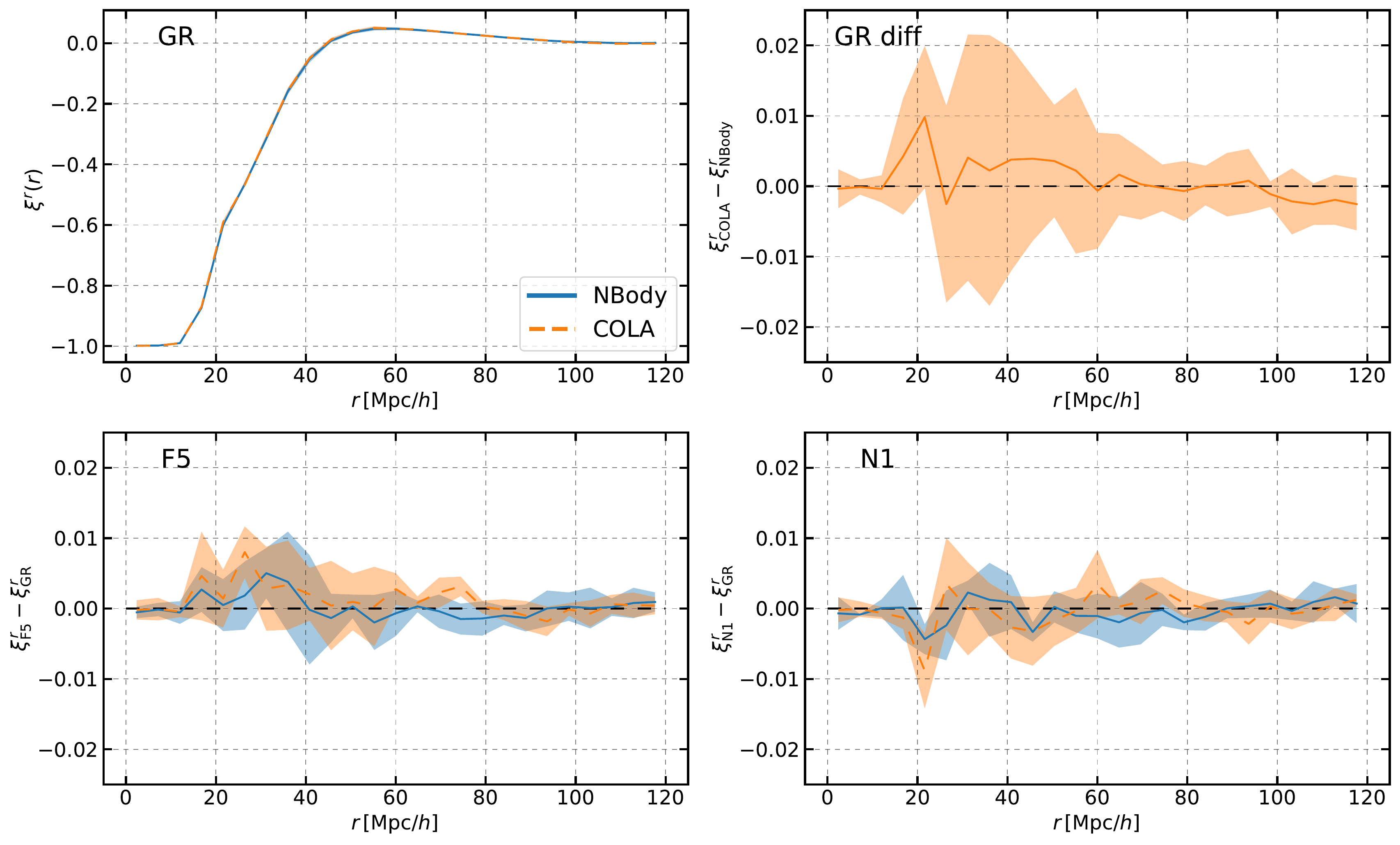}
\caption{\label{fig:CCF_real} Comparison of the galaxy-void real-space CCF in COLA (in orange) and {\it N}-body simulations (in blue). \textit{Top left:} Full signal in GR. \textit{Top right:} Difference of the GR signal between COLA and {\it N}-body simulations. \textit{Bottom:} Difference between the CCF in MG and GR for F5 and N1 gravity. The comparison uses differences instead of usual ratios since the signal crosses 0.}
\end{figure}

The real-space void-galaxy CCF in figure~\ref{fig:CCF_real} shows that COLA and {\it N}-body results are in agreement within the variance in all gravity models. Note that the top right panel shows the difference between COLA and {\it N}-body in GR instead of the usual ratio because the signal goes to zero on large scales. Similarly, the bottom panels show the differences between MG and GR results instead of the boost factors for the same reason. In both F5 and N1, there is no clear MG signal.

\begin{figure}
\centering
\includegraphics[width=.98\textwidth]{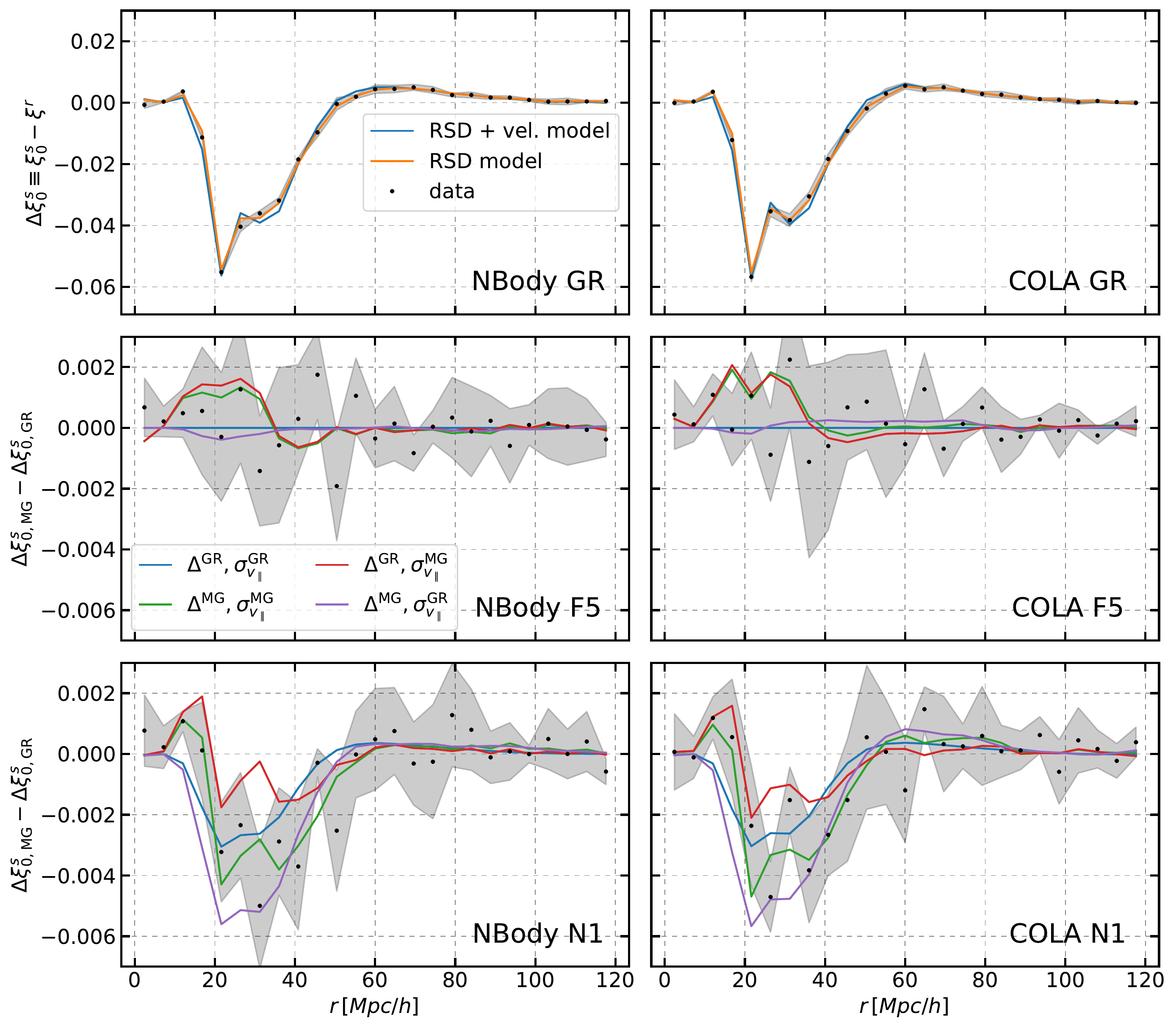}
\caption{\label{fig:xi0_COLAvsNBody_theory} Comparison of the monopole of the CCF of voids and galaxies in redshift-space between {\it N}-body (left column) and COLA (right column) simulations. \textit{Top panels:} The difference between the redshift-space monopole of the CCF and the real-space CCF, $\Delta\xi_{0}^s \equiv\xi_{0}^s-\xi^r$, measured in simulations (black dots and shaded region) is compared with the theory predictions (lines) formulated using the radial-velocity profile from simulations ("RSD model", orange) and using the velocity model of eq.~\eqref{VelModel} ("RSD + vel. model", blue). \textit{Middle and bottom panels:} The difference between $\Delta\xi_{0}^s$ in MG and GR (black dots and shaded regions) is compared to the theory predictions (lines) formulated using different combinations of inputs: integrated-density and velocity-dispersion from GR simulations (blue), integrated-density from MG and velocity-dispersion from GR simulations (violet), integrated-density from GR and velocity-dispersion from MG simulations (red), both integrated-density and velocity-dispersion from MG simulations (green). In all cases, we use the real-space CCF from GR simulations in the theory predictions, even in the MG panels to reduce the noise on the MG signals. The velocity model of eq.~\eqref{VelModel} is evaluated with the reference growth rate from linear theory (see table~\ref{tab:fBestFit}) in GR and N1 panels. In the case of F5, we use the reference growth rate from the GR linear theory.}
\end{figure}

\begin{figure}
\centering
\includegraphics[width=.98\textwidth]{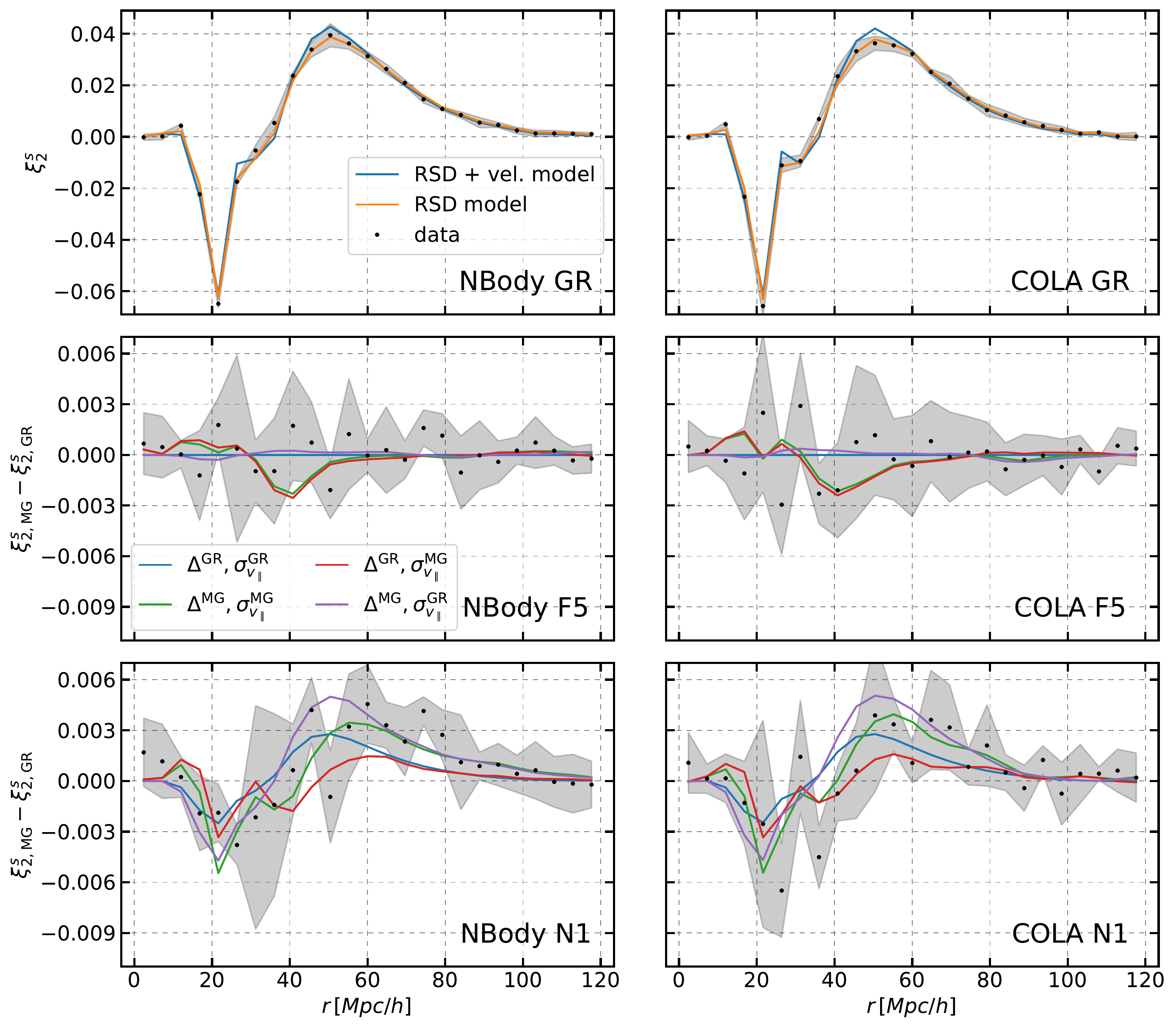}
\caption{\label{fig:xi2_COLAvsNBody_theory}Same as figure~\ref{fig:xi0_COLAvsNBody_theory} but for the quadrupole.}
\end{figure}

\begin{figure}
\centering
\includegraphics[width=.60\textwidth]{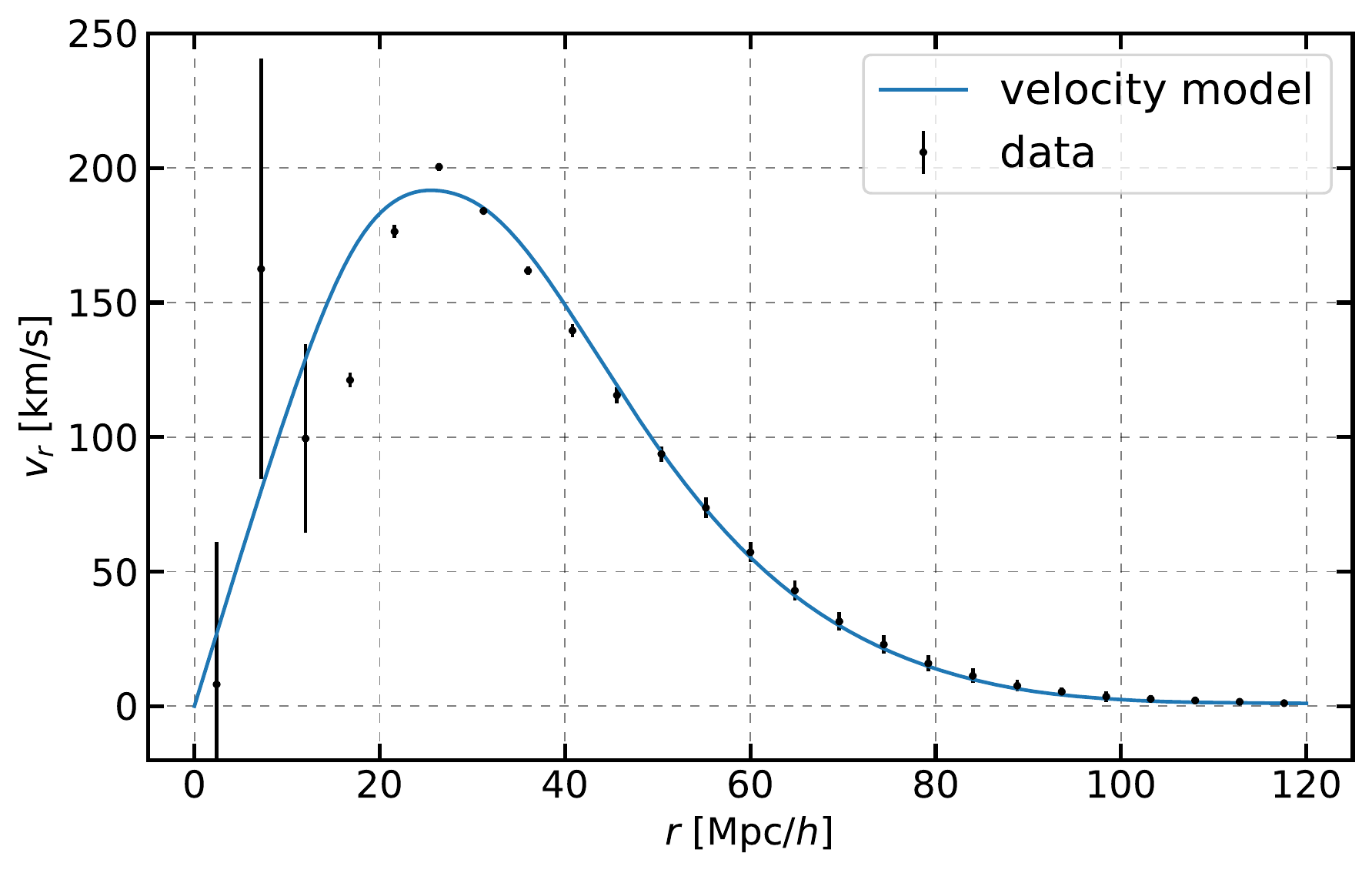}
\caption{\label{fig:DeltaVsVelProf} Comparison of the galaxy radial-velocity profile in GR {\it N}-body simulations with the linear model in eq.~\eqref{VelModel} using the fiducial growth rate $f=0.736$, as a function of the distance from the void centre.}
\end{figure}

In figure~\ref{fig:xi0_COLAvsNBody_theory} and figure~\ref{fig:xi2_COLAvsNBody_theory} we compare the monopole and the quadrupole of the void-galaxy CCF in redshift-space. In the top panels, the monopoles of the CCF ("data", black dots) are consistent within the variance (shaded regions) with the ones predicted using eq.~\eqref{theory_xiell} and the radial-velocity measured in the simulations ("RSD model", orange lines). This proves the accuracy of the RSD model discussed in section~\ref{ssec:RSDinVoids}. On the other hand, using the radial-velocity from eq.~\eqref{VelModel}%
\footnote{The value of the linear growth rate $f$ is taken from the linear theory.} 
in eq.~\eqref{theory_xiell} ("RSD + vel. model", blue lines) produces theoretical predictions that are still a good description of the data but with some discrepancies arising in the void interior that are traceable to the inaccuracies of the velocity model in reproducing the radial-velocity profile of galaxies in voids. 
The inaccuracies of the velocity model are visible in figure~\ref{fig:DeltaVsVelProf} where, comparing the velocity model with the radial-velocity profile estimated from {\it N}-body simulations in GR, we confirm the results of \cite{Nadathur:2017jos} that the model is a good description of the galaxy radial-velocity profile above $\sim 30 \mpcoh$ with some discrepancies arising below that distance from the void centre.

The bottom two rows of figure~\ref{fig:xi0_COLAvsNBody_theory} and figure~\ref{fig:xi2_COLAvsNBody_theory} show the impact of MG on monopole and quadrupole of the void-galaxy CCF in redshift space. In the case of F5, the signal is compatible with GR within the variance. Interestingly, N1 shows a clear enhancement of the distortions with respect to GR. This signal is well reproduced by using eq.~\eqref{theory_xiell} with the radial-velocity model computed in terms of $f^{\rm N1}$ from linear theory and $\Delta^{\rm N1}(r)$ and $\sigma_{v_\parallel}^{\rm N1}(r)$ estimated from N1 simulations (green lines). However, even ignoring the effects of MG on $\Delta(r)$ and $\sigma_{v_\parallel}(r)$ (i.e. using $\Delta^{\rm GR}(r)$ and $\sigma_{v_\parallel}^{\rm GR}(r)$ from GR simulations) but using the growth factor $f^{\rm N1}$ from N1 linear theory (blue lines) produces theoretical predictions that catch most of the signal (although representing a worse fit to the data). This is somehow surprising as we have seen from figure~\ref{fig:DeltaProfile} and figure~\ref{fig:VelDispProfile} that N1 gravity has impacts on both the integrated-density profile and the velocity-dispersion profile. To gain insight into this aspect, we add to the comparison the theoretical predictions formulated using the growth factor $f^{\rm N1}$ from N1 linear theory and, in one case $\Delta^{\rm GR}(r)$ and $\sigma_{v_\parallel}^{\rm N1}(r)$ (red lines), while in the other $\Delta^{\rm N1}(r)$ and $\sigma_{v_\parallel}^{\rm GR}(r)$ (violet lines). Comparing the two cases we can see that they have somehow competing effects on the redshift-space CCF and this explains why the main effect of N1 gravity is due to the enhancement of the linear growth factor.

For the monopole and quadrupole of the void-galaxy CCF in redshift-space (figure~\ref{fig:xi0_COLAvsNBody_theory} and figure~\ref{fig:xi2_COLAvsNBody_theory} respectively), results in COLA (right column) are consistent with the ones from {\it N}-body (left column) in all gravity models.

\subsection{Results for voids}
\label{ssec:VoidResults}
Using the RSD model discussed in subsection~\ref{ssec:RSDinVoids} with the velocity model of subsection~\ref{ssec:VelModel}, we leave the linear growth rate $f$ in eq.~\eqref{VelModel} as a free parameter and we test if we can recover the growth rate predictions from linear theory. 
We apply the delete-1 jackknife re-sampling technique \cite{Wu_Jackknife,Norberg:2008tg,2011MNRAS.418.2435N} to $n_{\rm sv}=125$ sub-volumes in which we split each realisation and average over the $n_{\rm r}=5$ realisations to estimate the covariance matrix for the multipoles of the CCF in redshift-space
\begin{equation}
    \matr{Cov}_{\ell, \ell', i,j} = \frac{1}{n_{\rm r}}\sum_{m=1}^{n_{\rm r}} \left[\frac{n_{\rm sv}-1}{n_{\rm sv}} \sum_{k=1}^{n_{\rm sv}} \left(\xi^s_{k}-\overline{\xi^s}\right)_{\ell,i}\left(\xi_{k}^{s}-\overline{\xi^s}\right)_{\ell',j} \right]_m \, ,
\end{equation}
where $\overline{\xi^s} = \frac{1}{n_{\rm sv}}\sum_{k=1}^{n_{\rm sv}} \xi^s_{k}$ is the average over the jackknife realisations of the CCF  $\xi^s_{k}$ estimated using 5 HOD realisations and 3 lines of sight. The indices $\ell, \ell'$ run over the multipoles $\{0,2\}$, while $i,j$ run over the radial bins. Given the periodic boundary conditions of the simulation box, we use the natural 2-point estimator
\begin{equation}\label{Natural2pEst}
    \widehat{\xi}_{\mathrm{N}}=\frac{D D}{R R}-1 \, .
\end{equation}

\begin{figure}
\centering
\includegraphics[width=.66\textwidth]{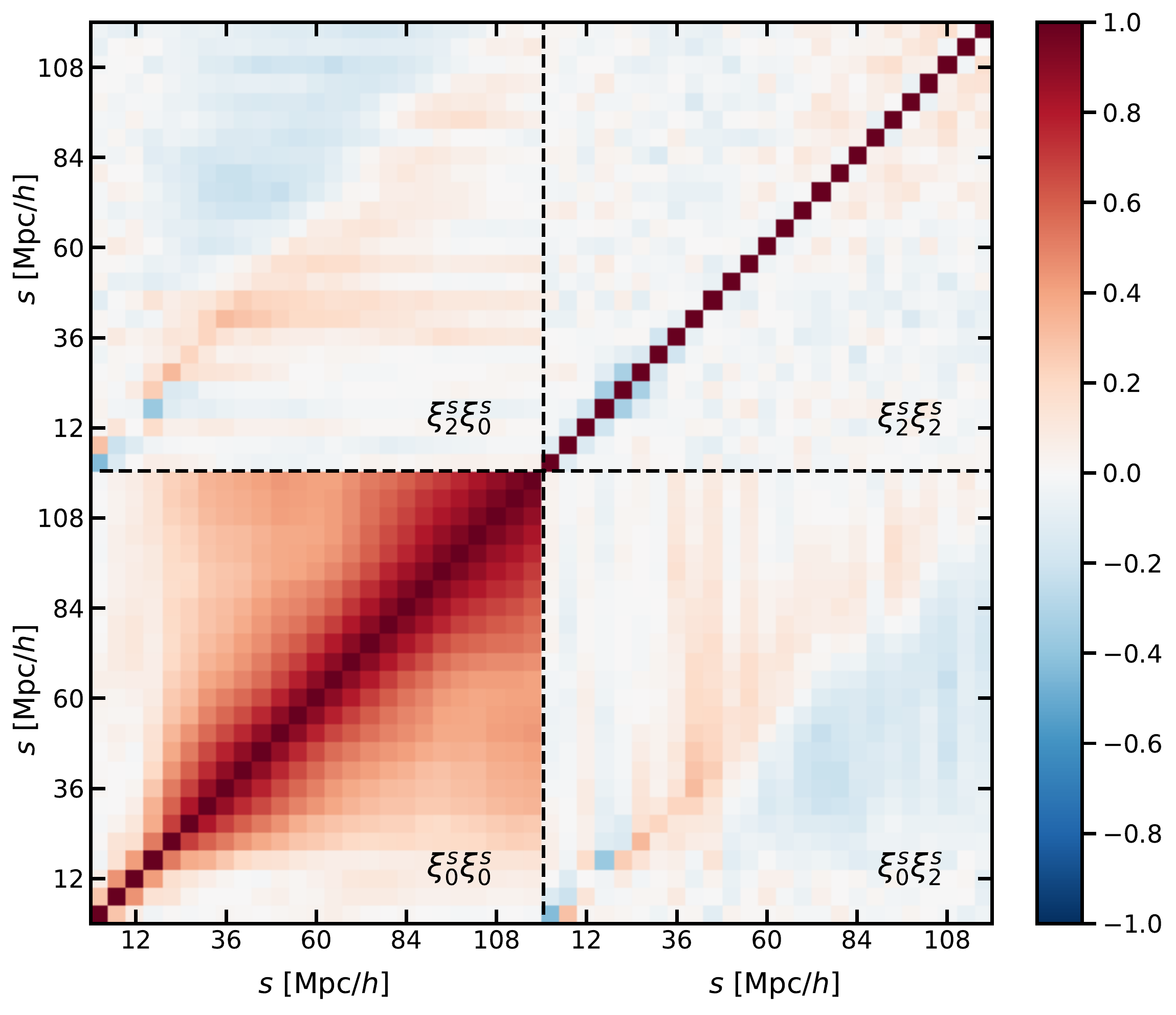}
\caption{\label{fig:CorrMat} Correlation of the covariance matrix estimated from jackknife re-sampling in {\it N}-body simulations in GR. The black dashed lines are used to distinguish between the monopole (bottom/left) and quadrupole (top/right) terms. The correlation matrices for the other gravity models as well as the ones for {\it N}-body simulations appear visually equivalent.}
\end{figure}

The correlation matrix obtained from the covariance matrix (figure~\ref{fig:CorrMat}) shows strong off-diagonal correlations between different bins in the monopole while the quadrupole is almost diagonal. From a visual comparison of our correlation matrix with the one in figure 6 of \cite{Nadathur:2017jos} we notice that in our case the off-diagonal correlations in the quadrupole are smaller. This is due to the fact that we average the signal over three lines of sight while \cite{Nadathur:2017jos} was using only 1 line of sight. Using only 1 line of sight likely produces a more realistic error estimate for a real survey but, since our goal in this work is to test the consistency of the RSD model in beyond-LCDM models and to validate approximate simulations, we choose to leverage on the average over three lines of sight (as well as the average over 5 HOD realisations) to increase the signal-to-noise ratio.

The recent work \cite{Mohammad:2021aqc}, has proposed the use of the Landy-Szalay 2-point estimator \cite{1993ApJ...412...64L}
\begin{equation}
    \widehat{\xi}_{\mathrm{LS}}=\frac{D D-2 D R+R R}{R R} \, ,
\end{equation}
to deal with the mask applied by construction by the jackknife re-sampling, as well as a correction to account for proper weighting of the "cross-pairs" (pairs with only one point falling in the sub-volume masked by the jackknife), which has been shown to produce more realistic covariance matrix estimates for the galaxy correlation function. In light of these results, we compute the jackknife covariance using the publicly available code \codeword{pycorr}\footnote{\href{https://github.com/cosmodesi/pycorr}{https://github.com/cosmodesi/pycorr}} and notice that:
\begin{itemize}
    \item Using the Landy-Szalay estimator produces smaller off-diagonal correlations in the covariance matrix for the monopole and larger errors on scales smaller than $\sim 20 \mpcoh$. The amplitude of the errors on these scales seems to depend on the number density of the random catalogues used (having tested for $20\times$ and $50\times$ the average number density of our galaxy and void catalogues) and shrinks for larger number density. This is compatible with shot-noise effects.
    \item Using the cross-pairs correction shrinks the errors on large scales in the monopole but does not significantly affect the correlations of the covariance matrix. 
\end{itemize}
Since this technique has not been tested yet against mocks in the specific case of the void-galaxy CCF, we make the conservative choice of adopting the same technique originally used to validate the RSD model in voids \cite{Nadathur:2017jos}, i.e. using the natural 2-point estimator in eq.~\eqref{Natural2pEst} and removing all the pairs involving an object falling in the masked sub-volume.

With the covariance obtained with these settings, we perform the fit for the linear growth rate $f$ by minimising the objective function
\begin{equation}\label{chi2}
    \chi^2= \sum_{\ell, \ell'}\sum_{i,j} (\xi^s_{\ell, i} - \xi^{\rm th}_{\ell, i}) \matr{Cov}_{\ell, \ell', i,j}^{-1} (\xi^s_{\ell', j} - \xi^{\rm th}_{\ell', j}) \, ,
\end{equation}
where $\xi^s$ is obtained by averaging over the 3 lines of sight, the 5 HOD realisations and the 5 realisations for each gravity model in COLA and {\it N}-body simulations. The theoretical predictions are based on $\xi^r$, $\Delta$ and $\sigma_{v_{\parallel}}$ estimated from the average of the signal over all the realisations in each gravity model in COLA and {\it N}-body simulations. 
We estimate the error on the best fit growth rate by finding the solutions to
\begin{equation}\label{sigmaf}
    \chi^2(f \pm \sigma_f) = \chi^2(f_{\rm fit})+1 \, .
\end{equation}
The results of the fit are shown in table~\ref{tab:fBestFit}. In GR, both COLA and {\it N}-body recover the reference linear growth rate within $\sim1$ standard deviation. 
\begin{table}[]
    \centering
        \begin{tabular}{cccccccc}
        \toprule
        {} & {} & \multicolumn{2}{c}{GR} & \multicolumn{2}{c}{F5} & \multicolumn{2}{c}{N1} \\
        {} & {} & COLA & {\it N}-body & COLA & {\it N}-body & COLA & {\it N}-body \\
        \midrule
        {} & $f_{\rm ref}$ & \multicolumn{2}{c}{0.736} & \multicolumn{2}{c}{--} & \multicolumn{2}{c}{0.777} \\
        \midrule
        {GR theory} & $f_{\rm fit}$ &  0.721 &  0.726 &  0.717 &  0.726 &  0.772 &  0.784 \\
        {} & $\sigma_f$ &  0.012 &  0.012 &  0.013 &  0.013 &  0.013 &  0.013 \\
        {} & $\Delta f /\sigma_f$ & -1.2 & -0.8 & -- & -- & -0.4 & 0.5 \\
        {} & $\chi^2/{\rm d.o.f.}$ & 0.96 & 1.08 & 0.88 & 0.92 & 1.53 & 1.69 \\
        \midrule
        {MG theory} & $f_{\rm fit}$ & -- & -- &  0.735 &  0.734 &  0.763 &  0.774 \\
        {} & $\sigma_f$ & -- & -- &  0.013 &  0.013 &  0.013 &  0.013 \\
        {} & $\Delta f /\sigma_f$ & -- & -- & -- & -- & -1.1 & -0.2 \\
        {} & $\chi^2/{\rm d.o.f.}$ & -- & -- & 0.67 & 0.77 & 0.78 & 0.95 \\
        \bottomrule
        \end{tabular}
    \caption{Results of the optimisation of eq.~\eqref{chi2} and eq.~\eqref{sigmaf} using theory predictions from GR simulations (top) and using theory predictions from MG simulations (bottom). COLA and {\it N}-body results are compared side by side in each gravity model (different columns). The additional row at the top of the table shows the linear theory predictions for the linear growth rate at redshift $z=0.5057$ in GR and N1. The linear theory growth rate in F5 is scale-dependent so we omit it from the table.}
    \label{tab:fBestFit}
\end{table}
In the case of N1 and F5 theories we investigate the impact of MG on the RSD model by performing the fit with different theoretical inputs:
\begin{itemize}
    \item using $(\xi^r$, $\Delta$, $\sigma_{v_{\parallel}})^{\rm GR}$ estimated in GR simulations  ("GR theory") ;
    \item using $(\xi^r$, $\Delta$, $\sigma_{v_{\parallel}})^{\rm MG}$ estimated in MG simulations  ("MG theory").
\end{itemize}
In both cases, we recover the reference growth rates within $\sim 1$ standard deviation in N1. In F5, we do not have a reference growth rate to compare our results with since the linear theory solution is scale-dependent, but the best fit values are consistent with the reference value in GR. Interestingly, comparing best-fit values obtained using "MG theory" and "GR theory", we notice a shift in the growth rate of $\sim 1 \sigma_f$ in both COLA and {\it N}-body simulations, towards larger values in the case of F5 and towards smaller values in the case of N1 using the GR theory. We also find an increase of the reduced $\chi^2$ in the GR theory. If this shift is in fact due to the effects of MG on the components entering the theoretical model, we might need to take into account these effects to get an accurate determination of the linear growth rate with $\sigma_f \lesssim 0.01$ precision with the void-galaxy CCF. At that level of precision, however, the inaccuracies of the velocity model of subsection~\ref{ssec:VelModel} (see also figure~\ref{fig:DeltaVsVelProf}) could become the limiting factor \cite{Massara:2022lng}. In such a scenario, simulation-based emulation of the void radial-velocity profile may be a way to avoid systematic errors of theoretical origin, and the COLA method could certainly be considered a viable option given its computational efficient nature and its accuracy in reproducing full {\it N}-body results as proven by this work. With less precision on the growth rate estimate, e.g. $\sigma_f \simeq 0.02$, voids analysis can still help constrain nDGP theories by providing independent information on the amplitude of the linear growth rate. In the particular case of N1, MG effects on the integrated-density profile and on the velocity-dispersion can be ignored in the light of the cancellation discussed in the previous subsection, but in general, this cancellation depends on the specific gravity model (and screening mechanism) and it needs to be carefully checked whether the same conclusion applies to other theories of gravity.

We performed the fit also using the covariance matrices corrected for the cross-pairs and/or computing the jackknife CCF with the Landy-Szalay estimator and confirmed that we obtained the same results.

\section{Conclusions} 
\label{sec:conclusion}
In this work, we used the simulation output of \cite{Fiorini:2021dzs} to investigate the effects of MG on 
the estimator for the real space power spectrum $Q_0$, bispectrum and voids while validating COLA simulations for these probes. 
In section \ref{sec:Q0} we compared the power spectrum orthogonal to the line-of-sight estimated by means of the truncated sum of eq.~\eqref{Q0ofP024}, $Q_0$, with the real space power spectrum $P^r$ in our halo and galaxy catalogues. 
For halos, we showed in figure~\ref{fig:Q0_halos} that $Q_0$ agreed within the variance with the real space power spectrum in the full range of scales considered and for all gravity models. By comparing COLA and {\it N}-body results, we found that the two were in agreement within the variance in GR and within $\sim 2 \%$ in the MG boost factors, the ratio between GR and MG.
For galaxies, we showed in figure~\ref{fig:Q0_galaxies} that the agreement between $Q_0$ and $P^r$ deteriorated after $k\sim 0.25 \hompc$ in COLA and after $k\sim 0.3 \hompc$ in {\it N}-body, meaning that higher multipoles need to be considered to recover $P^r$ at smaller scales. However, we found that the MG boost factors of $Q_0$ and $P^r$ were in excellent agreement at all scales, indicating that most of the MG information is contained in the first three even multipole moments. We determined that the accuracy of COLA is within the variance up to $k \sim 0.24 \hompc$ in GR and at all scales in the MG boost factors.

We examined the bispectrum of DM and galaxies in redshift space in section~\ref{sec:Bispectrum}, to understand how much of the MG bispectrum information filters through the HOD tuning. We compared results for bispectrum and reduced bispectrum to discriminate the part of the bispectrum signal due to the power spectrum from the one due to non-linearity. At the DM level, we showed in figure~\ref{fig:DM_bispec} that there were significant MG signals in the bispectrum, however, most of the MG signals in the bispectrum boost factors in F5 and N1 are due to the enhanced MG power spectrum as seen from the fact that the signal was much weaker in the reduced bispectrum. We found that COLA achieved $\sim 10\%$ accuracy in reproducing the {\it N}-body results in GR and $\sim 2\%$ in the N1 boost factor. In F5, we discovered that COLA simulations did not accurately reproduce the {\it N}-body bispectrum boost factors. We concluded that they lacked the contribution coming from non-linearity in the screening mechanism, which is neglected in the screening approximation used in COLA by linearising the scalar field equation. We highlighted this also by showing the configuration dependence of the matter bispectrum in figure~\ref{fig:DM_bispec_conf}, where COLA failed to reproduce the configuration dependence of {\it N}-body simulation results in F5, while it captured the {\it N}-body configuration dependence in N1 in which screening is weak on quasi non-linear scales. 
For galaxies in redshift-space, we have confirmed with  figure~\ref{fig:Galaxy_bispec_RS} that HOD tuning can hide the strong MG signal in the DM bispectrum coming from the power spectrum, with the resulting MG signals deviating only $\sim 5 \%$ from GR. By comparing COLA and {\it N}-body results, we found that COLA's accuracy is better than $10 \%$ in GR and $\sim 5\%$ in the MG boost factors. By looking at the configuration dependence of the boost factors for the bispectrum of galaxies in redshift space, we highlighted that this was dissimilar to the one observed in DM and concluded that the MG signal was dominated by non-linearity coming from galaxy bias. Due to this, the inaccuracy of COLA simulations in F5 for DM is largely hidden and COLA reproduced the configuration dependence of {\it N}-body simulations well.

In section~\ref{sec:Voids} we analysed the voids obtained from COLA and {\it N}-body galaxy catalogues with the \codeword{ZOBOV} void-finder. In order to test the applicability of the linear model of \cite{Nadathur:2017jos} to the case of MG, we carefully tracked the MG effects in each key component involved in the modelling. We showed in figures~\ref{fig:DeltaProfile},~\ref{fig:VelProfile} and~\ref{fig:VelDispProfile} that significant MG signals were present in the integrated-density, radial-velocity and velocity-dispersion profiles in N1, while in F5 only the velocity-dispersion profile showed a substantial enhancement compared to GR. We have not found any significant MG effect on the real space void-galaxy CCF, but by examining the monopole and quadrupole of the void-galaxy CCF in redshift space, we discovered that the RSD in N1 were significantly stronger than in GR, while in F5 they were not substantially different from the GR case. To understand the role of the linear growth rate, the integrated density and the velocity dispersion in the MG signal of the redshift-space void-galaxy CCF, we compared the predictions formulated using different combinations of theory components estimated from GR and MG simulations in  figure~\ref{fig:xi0_COLAvsNBody_theory} and figure~\ref{fig:xi2_COLAvsNBody_theory}. We concluded that the enhancements of the integrated-density profile and velocity-dispersion profile observed in N1 partially cancelled out and the linear growth rate played the leading role in enhancing the RSD in the void-galaxy CCF. Using the linear growth rate as a free parameter of the theory, we fitted the monopole and the quadrupole of the void-galaxy CCF in redshift space. We found that the best fit values were recovered within $\sim 1 \sigma$ of the reference value computed from the linear theory in GR and N1, even when using the integrated density and velocity dispersion from GR simulations in the theory predictions.
Concerning the accuracy of COLA in reproducing {\it N}-body results, we determined that COLA and {\it N}-body were consistent within the variance for all the void summary statistics analysed in GR and for the MG boost factors with the exception of the velocity-dispersion boost factor in N1 where COLA's accuracy is at $\sim 2\%$ level. We have found that the best fit values of the linear growth rate in COLA were consistent within $\sim 1\sigma$ with {\it N}-body. As for {\it N}-body simulations, it was possible to recover the linear growth rate reference values within $\sim 1\sigma$, even when integrated density and
velocity dispersion are taken from GR simulations.
We note that we used voids identified in real space galaxy mocks so that the real space void positions were known. When we use real observational data, they need to be recovered by applying a density-field reconstruction method \cite{Nadathur:2018pjn}. The reconstruction requires an assumption on the gravitational theory and there will be an additional MG effect in this process.

The use of multi-probes analysis will be key in breaking the degeneracies of the theory parameter space, enabling tight constraints of gravity on cosmological scales with next-generation experiments. The findings of this work are important to understanding the potential of $Q_0$, bispectrum and void-galaxy CCF in reducing the freedom of the galaxy bias model and in increasing the sensitivity to the gravity model. Furthermore, the validation of COLA simulations for these probes gives us a computationally cheaper alternative to full {\it N}-body simulations to formulate reliable theoretical predictions, bearing in mind the limitations of the approximations used in COLA simulations. This opens up the possibility to train beyond-$\Lambda$CDM cosmological emulators for these additional observables in sight of Stage IV surveys.

\acknowledgments
We thank Seshadri Nadathur and Slađana Radinović for useful insights and stimulating discussions on the void analysis. We thank Baojiu Li for providing the data from ELEPHANT simulations. 
For the purpose of open access, the authors have applied a Creative Commons Attribution (CC BY) licence to any Author Accepted Manuscript version arising from this work. AI and KK are supported by the UK Science and Technology Facilities Council (grant numbers ST/S000550/1 and ST/W001225/1).
Numerical computations were done on the Sciama High Performance Compute (HPC) cluster which is supported by the ICG, SEPNet and the University of Portsmouth. 

\paragraph{Data availability} Supporting research data are available on reasonable request from the corresponding author. 

\appendix 
\section{Bispectrum of halos and galaxies in real space}\label{sec:AppA}
For completeness, we add to the analysis of section~\ref{sec:Bispectrum} the results for the bispectra of halos and galaxy in real space.

\begin{figure}
\centering
\includegraphics[width=.98\textwidth]{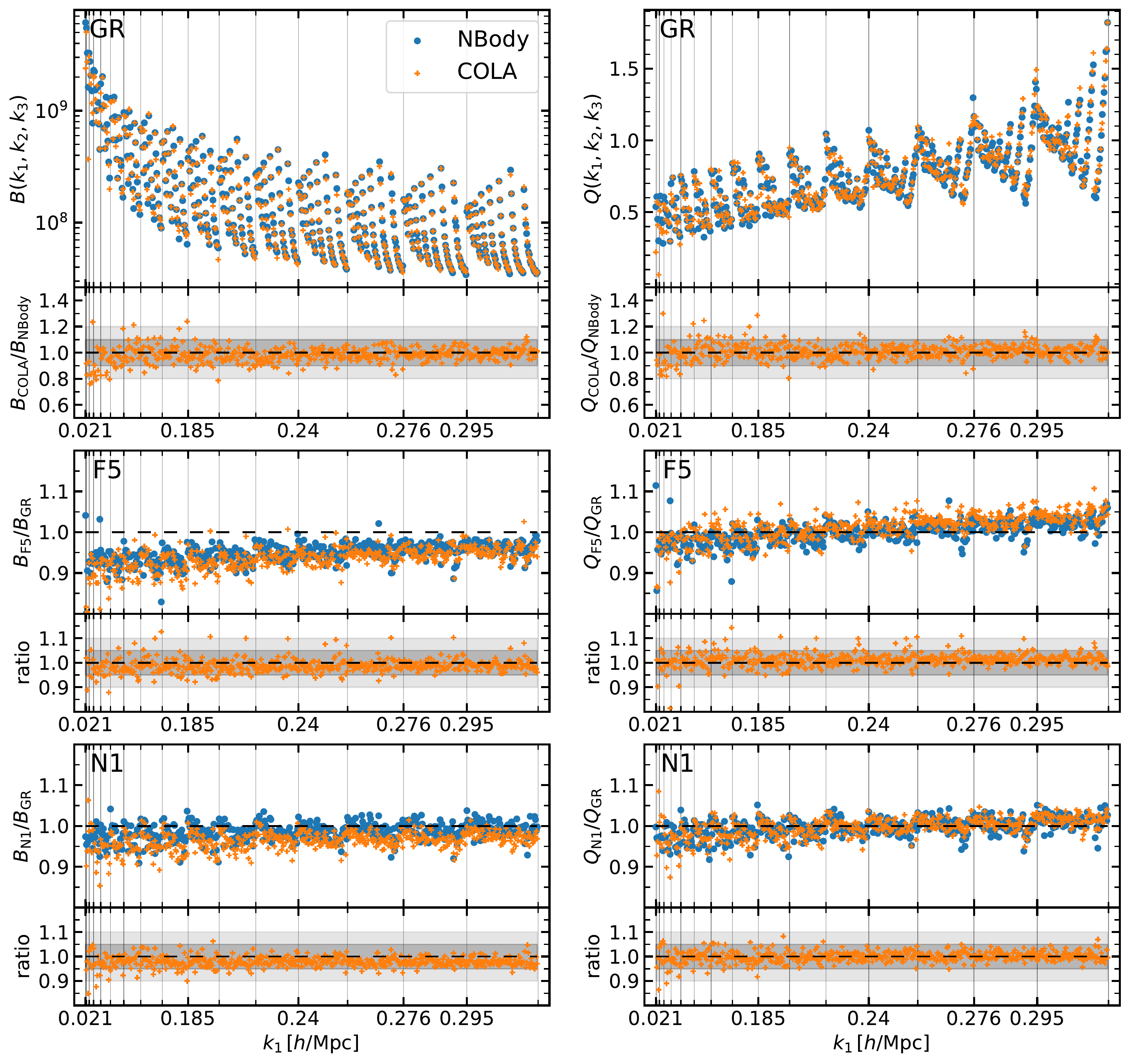}
\caption{\label{fig:Halo_bispec} Comparison of bispectrum (left column) and reduced bispectrum (right column) of halos in COLA (orange crosses) and {\it N}-body (blue dots) simulations. The configurations with $k_1 \ge k_2 \ge k_3$ are displayed in ascending order of the values of the wavenumbers $k_1$, $k_2$, and $k_3$ respectively. The vertical lines denote the value of $k_1$ for the configurations immediately to the right of each line. \textit{Top:} Full signal in GR. \textit{Middle and Bottom:} Boost factors in F5 and N1 respectively. In each panel, the bottom sub-panel shows the ratio between the COLA and {\it N}-body signals displayed in top sub-panel.}
\end{figure}

In the case of halo bispectrum, figure~\ref{fig:Halo_bispec} shows that COLA is able to reproduce the {\it N}-body signals for both bispectrum and reduced bispectrum of halos with $10\%$ accuracy in most of the configurations. The MG signal reaches $\sim 10\%$ in the case of the full bispectrum in F5, while it is limited to $\sim 5\%$ in the other cases. For the boost factors, COLA's accuracy is better than $\sim 5\%$ for most of the configurations.

The configuration dependence of the bispectra boost factors of halos is shown in figure~\ref{fig:Halo_bispec_conf}. Due to the bias of halos, the configuration dependence of DM in figure~\ref{fig:DM_bispec_conf} is not reflected in the configuration dependence of the halo bispectra boost factors. Thanks to this, the configuration dependence of COLA in F5 is qualitatively consistent with the one of {\it N}-body halo catalogues for both full and reduced bispectrum, despite the limitations in reproducing the matter bispectra discussed in section~\ref{sec:Bispectrum}.

The galaxy bispectrum in real-space is shown in figure~\ref{fig:Halo_bispec}. In GR, COLA is able to reproduce the {\it N}-body signals with $20\%$ accuracy for the full bispectrum and $10\%$ accuracy for the reduced bispectrum. The MG signals reach $\sim 10\%$ in the case of the full bispectrum in N1, while it is $\sim 5\%$ or less in the other cases. Despite the lower accuracy in reproducing the full bispectrum in GR, COLA's boost factors are $5\%$ accurate for most of the configurations. Figure~\ref{fig:Halo_bispec_conf} shows the configuration dependence of the bispectra boost factors of galaxies in real space. Again, the bias model dominates the MG signal hiding the clear MG boost factors of DM (figure~\ref{fig:DM_bispec_conf}). The strongest deviations from GR consistently appear towards the equilateral configuration (top right corner), in both full and reduced bispectrum independently of the gravity model. COLA tends to over-predict the MG signal of the bispectra in N1. Nonetheless, the qualitative description of the configuration dependence is consistent between COLA and {\it N}-body. 

\begin{figure}
\centering
\includegraphics[width=.98\textwidth]{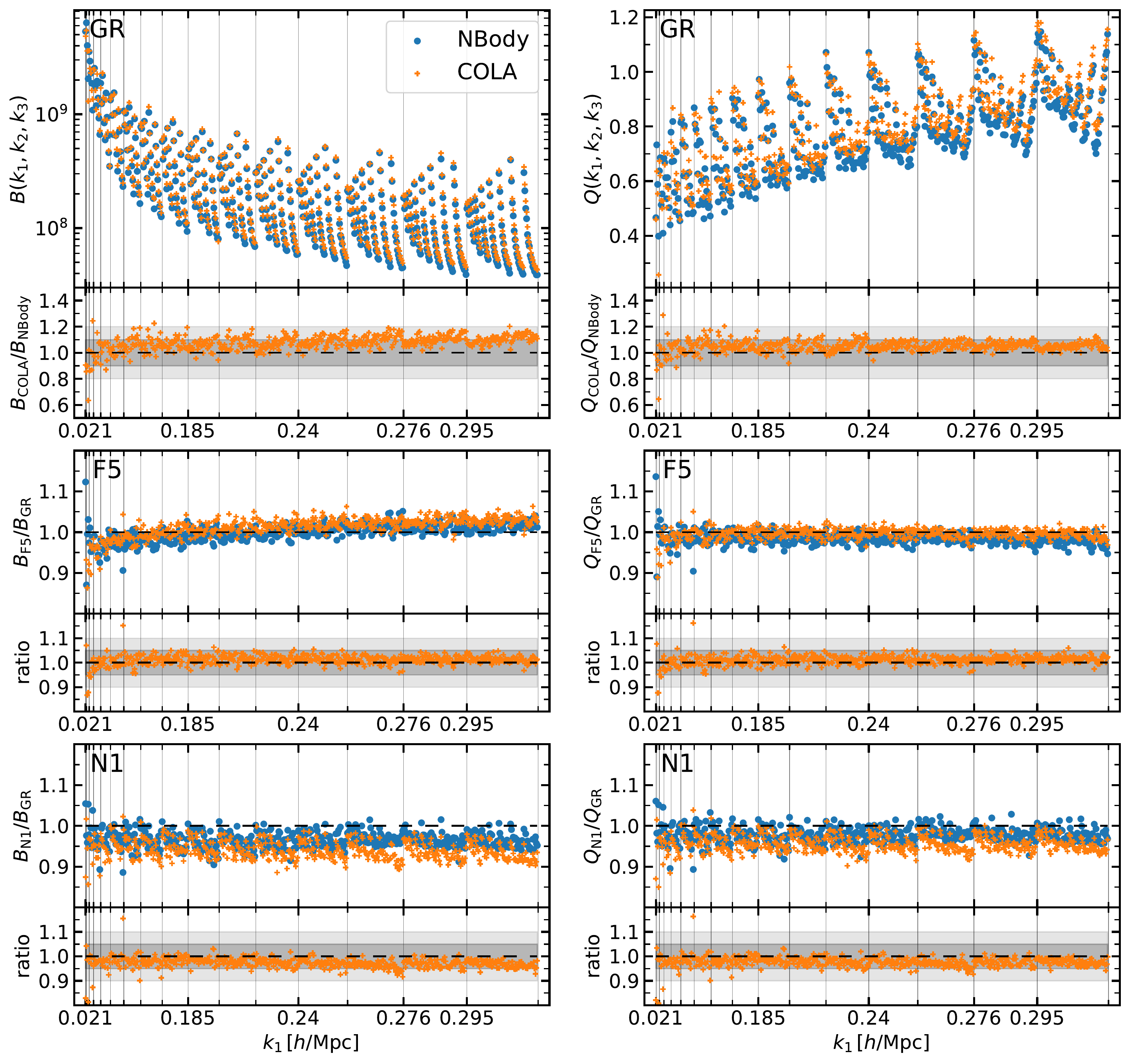}
\caption{\label{fig:Galaxy_bispec} Same as figure~\ref{fig:Halo_bispec} but for galaxies.}
\end{figure}

\begin{figure}
\centering
    \subfloat[][Full Bispectrum]{
    \includegraphics[width=.48\textwidth,clip]{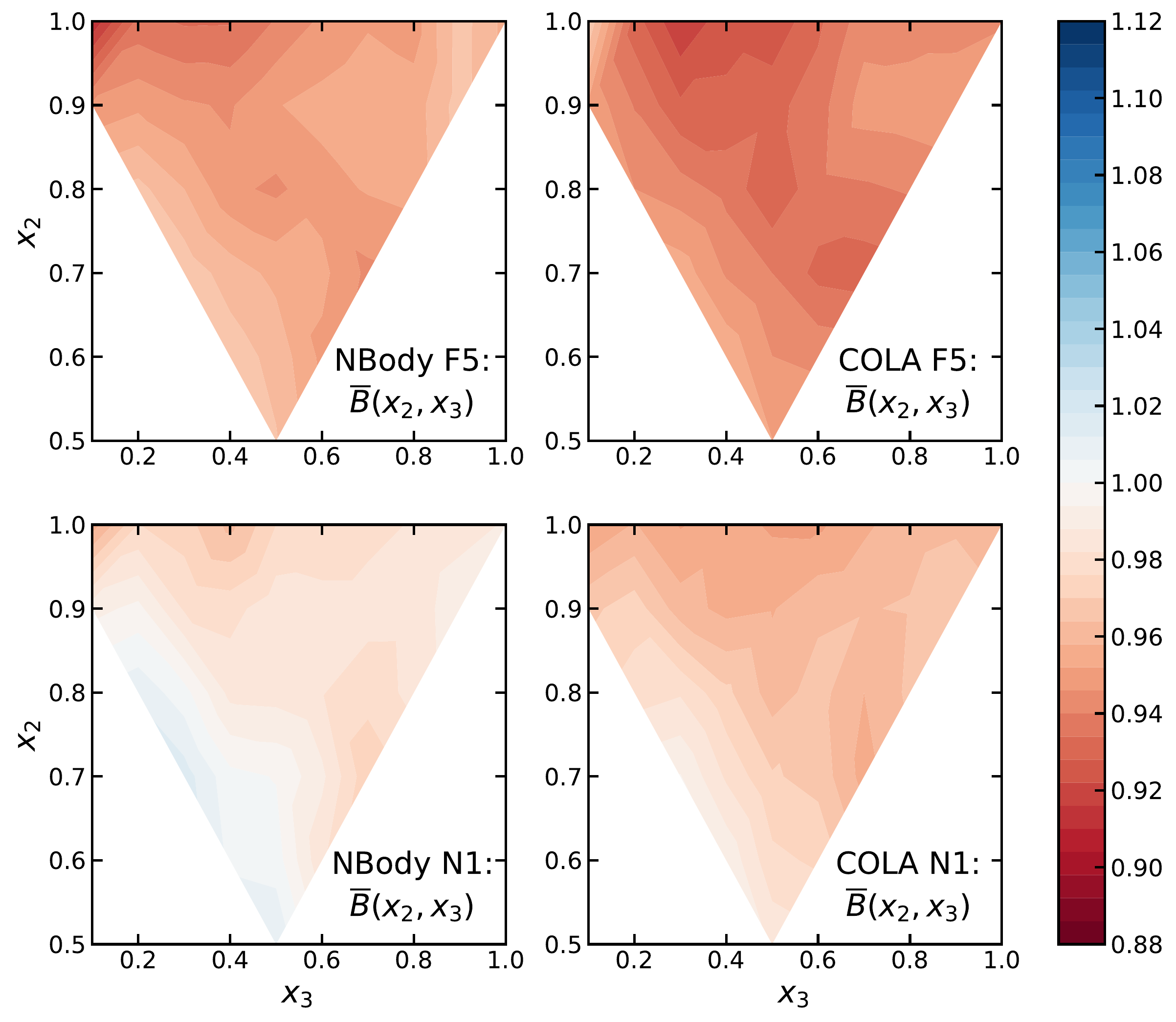}
    }
    \hfill
    \subfloat[][Reduced Bispectrum]{
    \includegraphics[width=.48\textwidth]{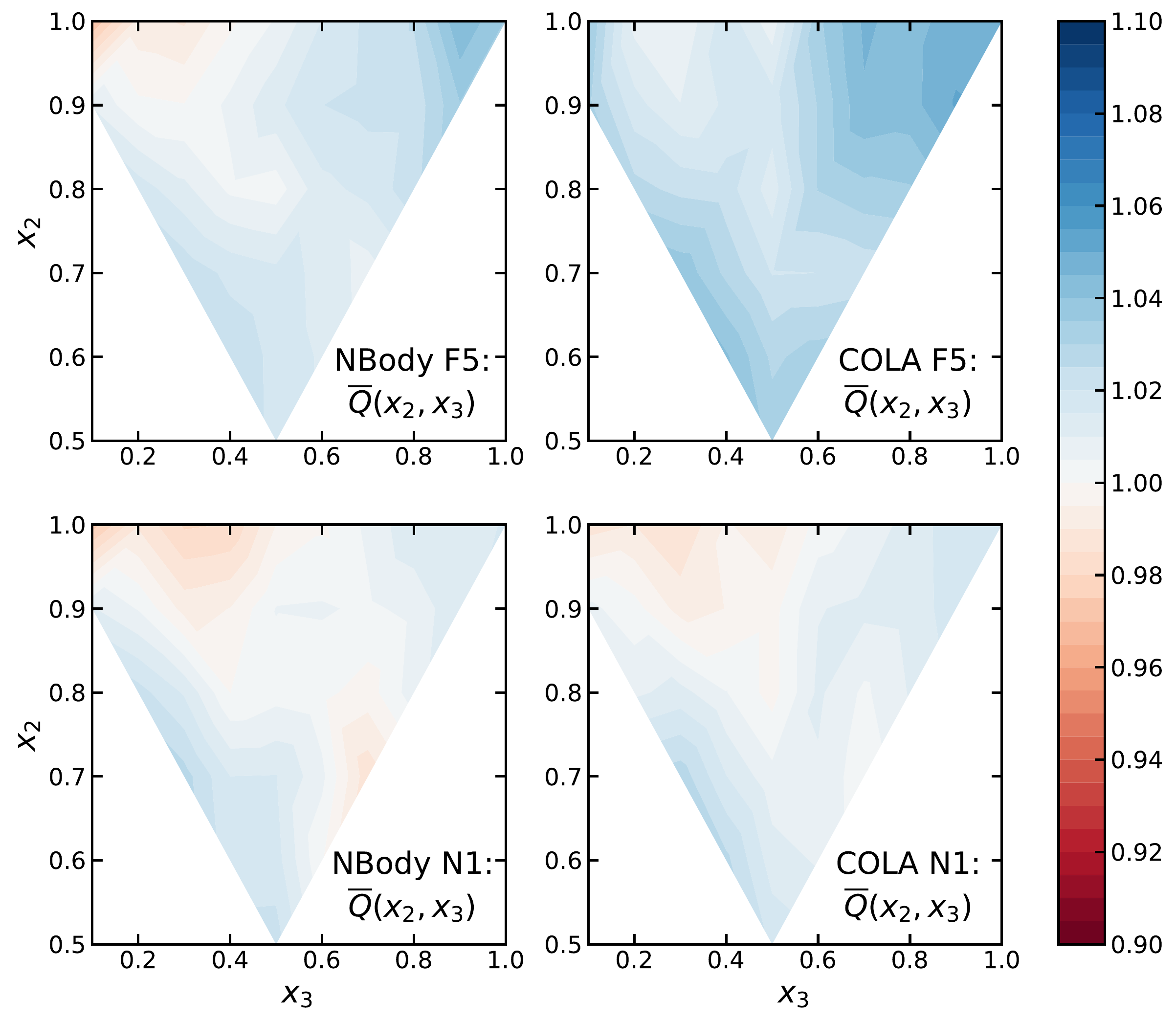}
    }
\caption{\label{fig:Halo_bispec_conf}Comparison of the configuration dependence of the F5 (top row) and N1 (bottom row) boost factors of bispectrum (on the left) and reduced-bispectrum (on the right) of halos in {\it N}-body (first and third columns) and COLA simulations (second and fourth columns). The colour-bars show the amplitude of the boost factors with blue (red) denoting stronger (weaker) signal in MG with respect to GR. \textit{In each panel:} The top right, top left and bottom corners of the triangle correspond to the equilateral, squeezed and folded configurations respectively. The squeezed configuration is missing from the figure due to the bin's width.}
\end{figure}

\begin{figure}
\centering
    \subfloat[][Full Bispectrum]{
    \includegraphics[width=.48\textwidth,clip]{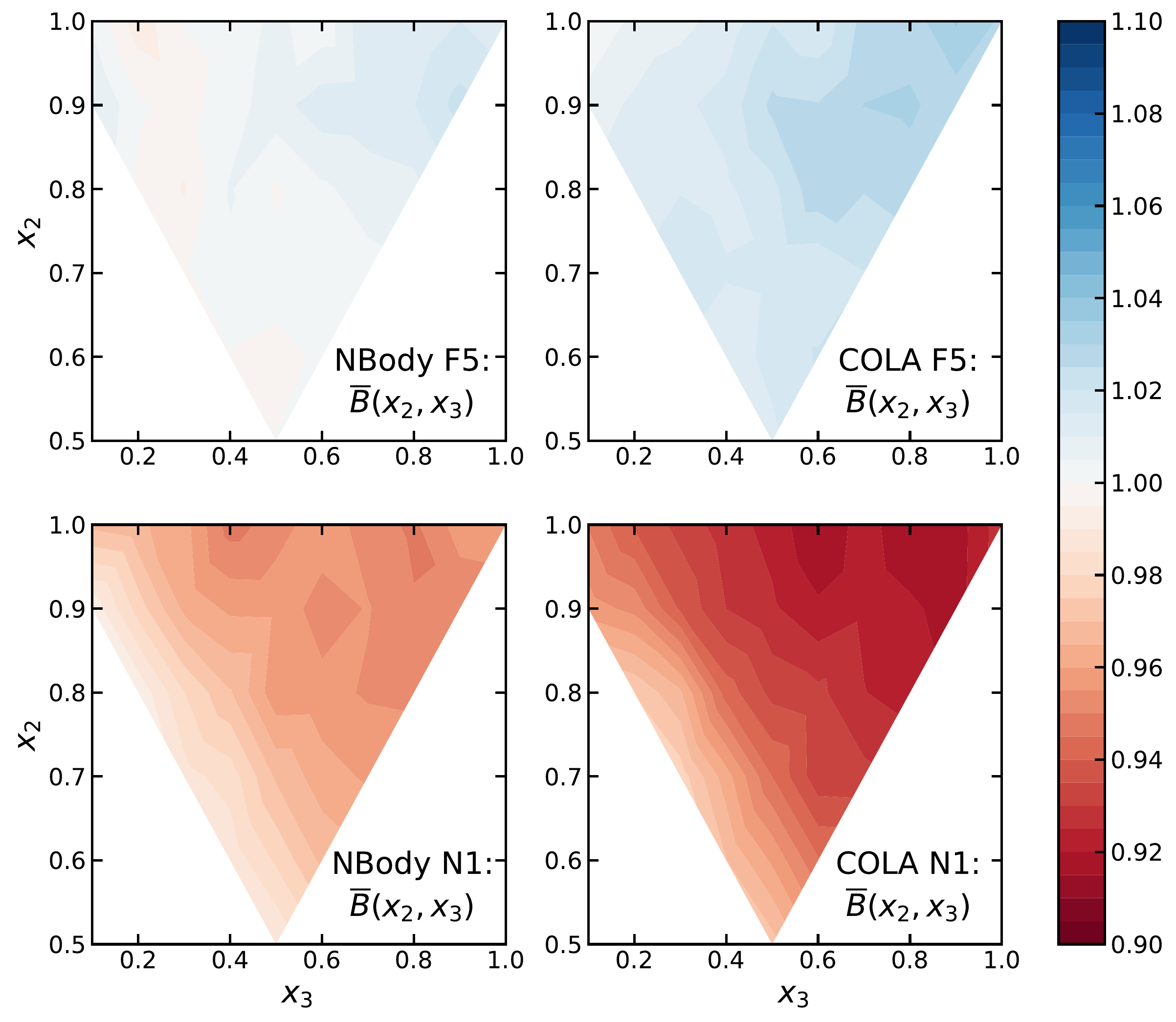}
    }
    \hfill
    \subfloat[][Reduced Bispectrum]{
    \includegraphics[width=.48\textwidth]{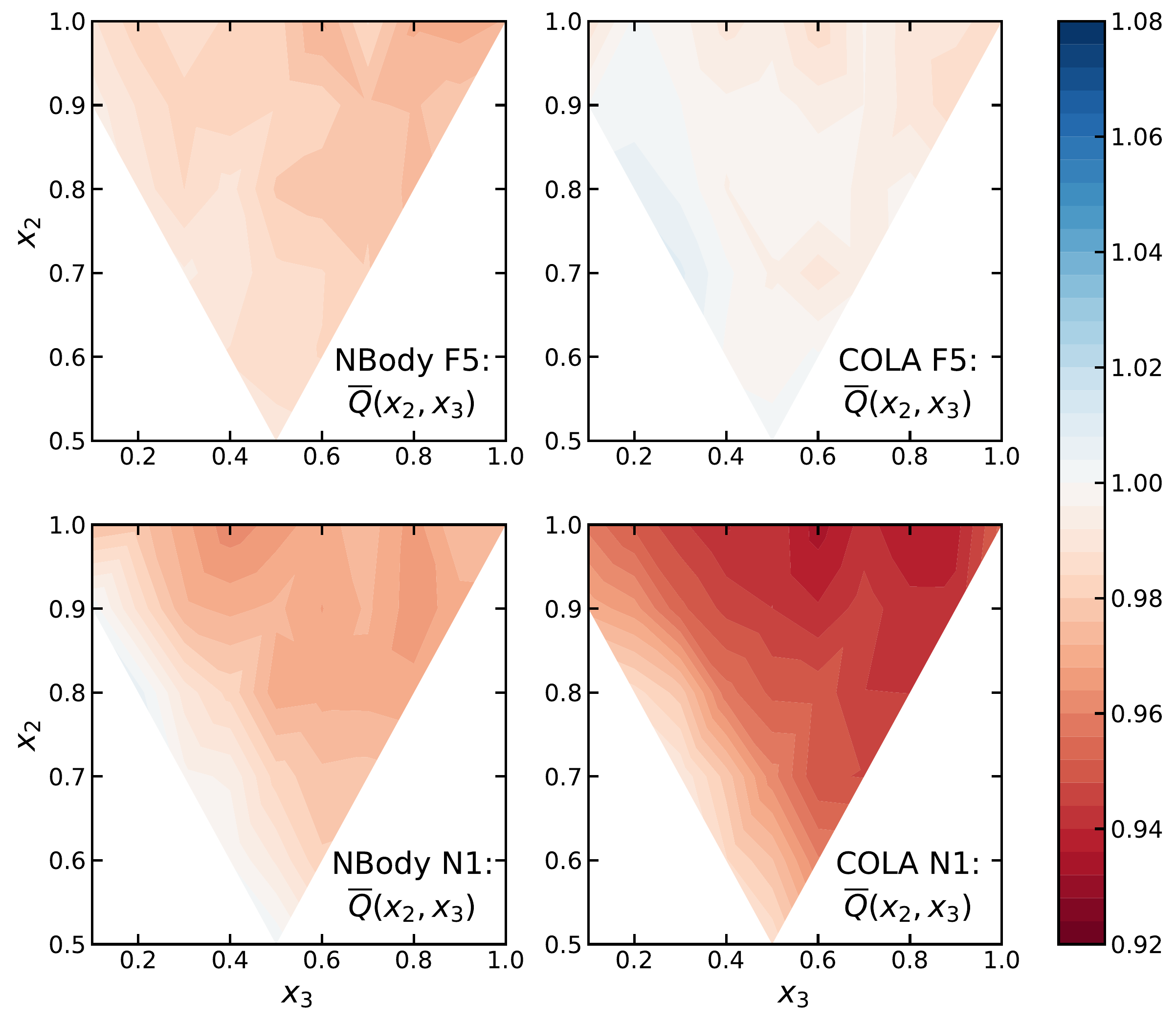}
    }
\caption{\label{fig:Galaxy_bispec_conf}Same as figure~\ref{fig:Halo_bispec_conf} but for galaxies.}
\end{figure}

\bibliographystyle{JHEP}
\bibliography{References}

\end{document}